\PassOptionsToPackage{colorlinks=true, urlcolor=blue, linkcolor=blue, citecolor=blue}{hyperref}
\documentclass[%
 reprint,
 amsfonts,amsmath,amssymb,
 aps,
 prd
]{revtex4-2}

\RequirePackage{graphicx}
\usepackage{amsmath, amsfonts, amsthm, mathrsfs, amssymb, color, amscd,
            xcolor, nicefrac, orcidlink}
\usepackage[%
  colorlinks=true,
  urlcolor=blue,
  linkcolor=blue,
  citecolor=blue
]{hyperref}

\newcommand{\beq}{\begin{equation}}
\newcommand{\eeq}{\end{equation}}
\newcommand{\bea}{\begin{eqnarray}}
\newcommand{\eea}{\end{eqnarray}}

\providecommand{\dif}{\mathrm{d}} 

\usepackage[normalem]{ulem}                    

\def\DD{\Delta}

\def\Opava{Research Centre for Theoretical Physics and Astrophysics,
Institute of Physics, Silesian University in Opava, CZ-74601 Opava,
Czech Republic}

\begin{document}


\title{Charging of rotating black holes: kinetic simulations of black hole magnetospheres}

\def\Opava{Research Centre for Theoretical Physics and Astrophysics, Institute of Physics, Silesian University in Opava, Bezru\v{c}ovo n\'am.~13, CZ-74601 Opava, Czech Republic}

\def\MPIfR{Max-Planck-Institut f\"ur Radioastronomie,
  Auf dem H\"ugel 69, D-53121 Bonn, Germany}

\def\FZU{Institute of Physics of the Czech Academy of Sciences, Na Slovance 1999/2, Prague 8, Czech Republic}

\def\ICRANet{ICRANet, Piazza della Repubblica 10,
  I-65122 Pescara, Italy}
  
\def\ICRA{ICRA, Dipartimento di Fisica, Sapienza Universit\`a di Roma,
  Piazzale Aldo Moro 5, I-00185 Rome, Italy}

\def\icranetferrara{ICRANet-Ferrara, Dipartimento di Fisica e Scienze della Terra, Universit\`a degli Studi di Ferrara, Via Saragat 1, 44122 Ferrara, Italy}

\def\unife{Dipartimento di Fisica e Scienze della Terra, Universit\`a degli Studi di Ferrara, Via Saragat 1, 44122 Ferrara, Italy}

\def\iaps{INAF, Istituto di Astrofisica e Planetologia Spaziali, Via Fosso del Cavaliere 100, I-00133 Rome, Italy}
  
\def\IPAG{Univ. Grenoble Alpes, CNRS, IPAG,
  F-38000 Grenoble, France}


\author{Martin Kolo\v{s}\;\orcidlink{0000-0002-4900-5537}}
  \email{martin.kolos@physics.slu.cz}
  \affiliation{\Opava}

\author{Farukh Abdulkhamidov\;\orcidlink{0000-0001-6424-9516}}
  \email{farukh.abdulkhamido@physics.slu.cz}
  \affiliation{\Opava}

\author{Arman Tursunov\;\orcidlink{0000-0001-5845-5487}}
  \email{tursunov@fzu.cz}
  \affiliation{\FZU}
  \affiliation{\Opava}

\author{Jorge A. Rueda\;\orcidlink{0000-0003-4904-0014}}
  \email{jorge.rueda@icra.it}
  \affiliation{\ICRANet}
  \affiliation{\ICRA}
  \affiliation{\icranetferrara}
  \affiliation{\unife}
  \affiliation{\iaps}

\author{Beno\^{i}t Cerutti\;\orcidlink{0000-0001-6295-596X}}
  \email{benoit.cerutti@univ-grenoble-alpes.fr}
  \affiliation{\IPAG}

\begin{abstract}
Whether a rotating black hole (BH) immersed in an external magnetic field charges up to the Wald value $Q_{\rm W}=2aMB$, where $M$ and $a$ are the BH mass and spin parameter, and $B$ is the strength of the external field, is a long-standing open question in BH electrodynamics, with consequences for charge separation, particle acceleration and the structure of BH magnetospheres. We address it with axisymmetric general-relativistic particle-in-cell simulations performed with GRZeltron, including self-consistent pair creation. For representative values of the BH spin, we evolve the same asymptotically uniform Wald field from two opposite initial horizon charges, $Q_{0}=0$ and $Q_{0}=Q_{\rm W}$, and track the accumulated charge through the BH horizon. We find that both branches relax within a few tens of gravitational times to the same saturated charge value. Thus, the equilibrium is a dynamical attractor of the kinetic magnetosphere rather than a memory of the initial data. The attractor lies well below $Q_{\rm W}$, at $\xi_{\rm eq}\equiv Q_{\rm eq}/Q_{\rm W}\approx 0.3$ for $a\lesssim0.7$, then falling almost to $\xi_{\rm eq}\approx0$ for large spins $a\approx1$. 
We derive an analytic expression for $\xi_{\rm eq}$ by requiring equal magnetic fluxes through the horizon associated with positive and negative charges, and find its spin dependence to be in agreement with the simulations.
The charge state of an astrophysical BH is therefore driven by kinetic plasma processes, and the Wald charge is an upper bound, not a general equilibrium value. Since $\xi_{\rm eq}<1$, charging does not quench the horizon-infinity potential drop that powers energy-extraction processes such as the Blandford-Znajek mechanism.
\end{abstract}


\maketitle

\section{Introduction}\label{sec:intro}

Astrophysical black holes (BHs) are never isolated objects. They are embedded in the magnetized, collisionless plasma of an accretion flow, and the magnetic field carried by that plasma threads the horizon and extends far beyond it. Millimetre-VLBI observations of M87$^{*}$ and Sgr~A$^{*}$ have now resolved this environment on horizon scales \cite{EHT:2019:ApJL:,EHT:2022:ApJL:}, and the high linear polarization of the emission indicates dynamically important, ordered poloidal fields consistent with a magnetically arrested accretion state \cite{EHT:2021:ApJL:}. The field strengths inferred at horizon scales range from tens of gauss in low-luminosity active galactic nuclei to $10^{4}$--$10^{8}$~G in more compact or more luminous systems. In these systems, the magnetized BH, not the isolated Kerr BH, is the relevant object of study.

In astrophysical studies of such an environment, the BH is traditionally assumed to be neutral. The usual justification is that the ambient plasma would discharge the BH charge almost instantaneously in  
a light-crossing time. This argument, however, does not apply when the system's electrodynamics continuously regenerates the charge. A rotating BH in an external magnetic field induces, through unipolar induction, an electric field that cannot be screened by the geometry alone. This field separates the surrounding plasma charges, causing the horizon to absorb opposite-sign charges at different rates. The natural end state of this process is not $Q=0$, but a nonzero equilibrium charge at which the selective accretion switches off \cite{Wald:1974:PHYSR4:,Tur-Stu-Kol:2016:PRD:,Zaj-etal:2018:MNRAS:,Tur-etal:2020:ApJ:}.

The reference starting point for all subsequent discussion is the stationary, electro-vacuum solution of the Einstein-Maxwell equations for a Kerr BH in an asymptotically uniform (test) magnetic field, aligned with the BH spin axis, the Wald solution \cite{Wald:1974:PHYSR4:}. Using this solution, Wald argued that the BH evolves through a selective charge accretion process that stops when the charge reaches an equilibrium value
\beq
Q_{\rm W} = 2 B J = 2 a M B,
\label{eq:qwald}
\eeq
at which the electric potential difference between spatial infinity and the horizon, on the symmetry axis, vanishes. Here, $M$ is the mass of the BH, $J = a M$ its angular momentum ($a$ is the angular momentum per unit mass or spin parameter), and $B$ is the strength of the asymptotically uniform magnetic field. Hereafter, we refer to $Q_{\rm W}$ as the Wald charge. For our discussions, it is convenient to normalize the BH charge by this value and define
\beq
\xi \equiv \frac{Q}{Q_{\rm W}}.
\label{eq:xi}
\eeq
The central question is whether an astrophysical BH, embedded in a plasma-filled magnetosphere, evolves toward the Wald equilibrium charge, $\xi_{\rm eq} = \xi_W = 1$, or obeys a different value. 

This question has remained open because different theoretical approaches have led to qualitatively distinct expectations. As we recalled above, in the original electro-vacuum picture, $Q_{\rm W}$ is the preferred equilibrium charge state. Alternative global arguments, however, do not support $\xi_{\rm eq} = 1$ as a unique equilibrium. In \cite{LiXin:2000:PRD:}, it was found that the electromagnetic energy is minimized at a sub-Wald charge $Q = \xi(a)Q_{\rm W}$, where $\xi(a)$
reaches a maximum of $\approx 0.13$ for extremal spin ($a = 1$), and vanishes rapidly as $a$ decreases. More recently, selective-accretion models have revisited the problem from the viewpoint of charge-dependent absorption. In the charged test-particle analysis \cite{Rue-Ruf:2024:EPJC:}, the charged Wald electromagnetic field divides the horizon into polar and equatorial sectors that preferentially accrete opposite charge signs, with a separation angle depending sensitively on $\xi$. Based on this result, it was argued that the BH should enter an oscillating behavior around a certain charge $\xi <1$. In a dilute-accretion framework \cite{Ber-etal:2025:arXiv:}, it has been shown that $Q_{\rm W}$ is not a universal exact saturation charge, although it may remain the leading-order result in the strong-field dilute limit. Conversely, a sharp restatement of the Wald-screening picture was given in \cite{Kin-Pri:2021:ApJL:}, where it was argued that a rotating BH in an aligned magnetic field accretes toward $\xi \approx 1$, thereby suppressing the electric field that drives energy extraction via the
Blandford-Znajek mechanism. 

A second family of descriptions replaces the vacuum with an idealized conducting magnetosphere. However, force-free electrodynamics approaches ensure that $\vec{E}\cdot\vec{B} = 0$ holds everywhere, so by construction they cannot address the present problem, as it removes the parallel electric field that separates charges. A related subtlety concerns the near-extreme spin limit: in vacuum, the magnetic flux is completely expelled from the horizon as $a \to M$, the BH analog of the Meissner effect \cite{Bic-Jan:1985:MNRAS:}, but simulations of conductive magnetospheres show that the currents flowing in the plasma largely suppress this expulsion \cite{Kom-McK:2007:MNRAS:}. Whether the charging mechanism shuts off in the extreme BH limit is therefore not obvious a priori.

The plasma problem is even more complex. In a plasma-filled magnetosphere, global electric currents, pair creation that supplies the plasma, and the overall electrodynamic structure determine the electric field. A key step in this direction was taken in \cite{Kom:2004:MNRAS:}, where the magnetospheric Wald problem was studied in resistive electrodynamics, and it was shown that an equatorial current sheet develops inside the ergosphere, where magnetic field lines are forced to rotate. Later, in a re-analysis of the charged Wald solution, it was shown in  \cite{Kom:2022:MNRAS:} that even at $Q = Q_{\rm W}$, the electrostatic potential drop along most magnetic field lines does not vanish, implying a live magnetosphere. Thus, the equilibrium charge would satisfy $\xi < 1$ with its value depending on how charged particles are supplied to the magnetosphere.

First-principles kinetic simulations are therefore a natural way to address the problem, since they make no assumption about the plasma supply, the conductivity, or the degree of screening. The first general-relativistic particle-in-cell (GRPIC) studies of BH magnetospheres treated the one-dimensional problem of a charge-starved gap above the horizon, in which pair cascades fed by the accretion-disk photon field
intermittently screen the parallel electric field \cite{Lev-Cer:2018:AA:}. 
A vacuum magnetosphere and one populated by free charges are fundamentally different objects, and the difference was demonstrated directly by global two-dimensional calculations \cite{Par-Phi-Cer:2019:PRL:}. There, the vacuum Wald solution of the Maxwell equations served only as the initial condition, while the magnetosphere was supplied with $e^+e^-$ pairs through a volumetric injection controlled by the local value of $\vec{E}\cdot\vec{B}$. Once free charges are present, the initial vacuum configuration is rapidly abandoned: the injected plasma screens the parallel electric field, the magnetic field lines bend back toward the BH, an equatorial current sheet forms within the ergosphere, and the system converges towards a current-carrying, Blandford-Znajek-like magnetosphere that powers a jet. Radiative GRPIC simulations of pair discharges likewise show that the gap opens near the inner light surface and intermittently injects pair plasma into the magnetosphere, while both charge species can fall through the horizon inside the inner light surface \cite{Cri-etal:2020:PRL:,Cri-etal:2021:AA:}. Comparable results have been obtained with independent codes \cite{Che-Yua:2020:ApJ:,Bra-Rip-Phi:2021:PRL:}. The problem has since been extended to 3D and to synthetic images of the reconnecting equatorial sheet \cite{Cri-etal:2022:PRL:}, and to magnetospheres magnetically connected to a surrounding disk \cite{ElM-etal:2022:AA:}. These results strongly suggest that the charge acquired by the BH is a dynamical kinetic quantity, controlled by the competition between horizon absorption of opposite charge signs in a pair-producing plasma. To our knowledge, however, none of these studies has tracked the net charge accumulated by the BH itself, which is precisely the quantity at issue in the Wald
problem.

This motivates the present work. We study the charging of rotating BHs in axisymmetric GRPIC simulations performed with the GR version of the \textsc{Zeltron} code \cite{Cer-Wer:2019:ascl:}. Each simulation begins from the same asymptotically uniform Wald field at fixed spin, but we consider two distinct initial conditions for the BH charge: $Q_{0} = 0$ and $Q_{0} = Q_{\rm W}$. The simulations include pair production, radiation reaction, and the full kinetic plasma response. We measure the charge accumulated by the BH from the Gauss-law flux of the electric field through a surface enclosing the horizon.
This setup allows us to address two questions. First, does the system retain a memory of the initial BH charge, or do both branches converge to the same asymptotic value of $\xi$? Second, if they do converge, how far does the resulting equilibrium deviate from the Wald limit, and how does this deviation depend on the spin?
The answer to the first question, established in Sec.~\ref{sec:results}, is that both branches converge. For every spin considered, the runs started from $Q_{0} = 0$ and from $Q_{0} = Q_{\rm W}$ converge to the same positive, sub-Wald charge. Thus, the horizon charge behaves as a dynamical attractor without retaining memory of the initial condition. The attractor lies at $\xi \approx 0.3$, favoring the plasma-regulated picture of \cite{Kom:2004:MNRAS:,Kom:2022:MNRAS:} and the test particle conjecture \cite{Rue-Ruf:2024:EPJC:} over the Wald value \cite{Wald:1974:PHYSR4:} and the much smaller minimum-energy charge of \cite{LiXin:2000:PRD:}. This provides, to the best of our knowledge, the first direct GRPIC determination of the equilibrium BH charge in the Wald problem.

The paper is organized as follows. Section~\ref{sec:mechanism} develops the analytic charging mechanism: the charged Wald solution, the horizon-infinity potential difference, the separatrix $\vec{E}\cdot\vec{B} = 0$, and the equilibrium condition that follows from balancing the magnetic flux subtended by the two charge-capture sectors of the horizon. Section~\ref{sec:grpic} describes the GRPIC method, the GRZeltron code, the numerical setup, and the diagnostics used to measure the horizon charge. Section~\ref{sec:results} presents the measured equilibrium charge as a function of spin for both initial conditions and compares it with the analytic prediction. Section~\ref{sec:summary} summarizes our conclusions and discusses the astrophysical implications and the limitations of the axisymmetric treatment. Throughout, we use the Kerr metric and, without loss of generality, set $G = c = 1$ units.

\section{Black hole charging mechanism}\label{sec:mechanism}

The BH charge is gravitationally negligible when it is much smaller than the characteristic value
\beq
Q_{\rm G} \sim \sqrt{G}\,M \approx
10^{30}\left(\frac{M}{M_\odot}\right)\,{\rm statC},
\label{eq:QG}
\eeq
at which its contribution to the spacetime curvature is comparable to that of the mass. An induced charge on a rotating BH, of the order of the Wald charge
\beq
Q_{\rm W} \sim 2aMB \lesssim
10^{18}\left(\frac{M}{M_\odot}\right)^{2}
\left(\frac{B}{10^{8}\,{\rm G}}\right)\,{\rm statC},
\label{eq:QWphys}
\eeq
lies many orders of magnitude below this limit. Therefore, we can treat both the induced charge and the external magnetic field as test quantities that do not modify the background geometry.

The smallness of $Q/Q_{\rm G}$ should not,  however, be mistaken for dynamical irrelevance, as what enters in the equation of motion of a plasma particle is not $Q/M$ but $qQ/(mM)$. The electron specific charge $q/m$ is larger by some twenty-one orders of magnitude than the value at which electromagnetic and gravitational forces would be comparable. Thus, a BH charge far too small to bend geodesics can completely reorganize the trajectories of the plasma that supply the magnetosphere.

At present, no universal model exists for the electromagnetic fields around astrophysical BHs, and a variety of theoretical frameworks have
been proposed
\cite{Bic-Jan:1985:MNRAS:,Kom:2004:MNRAS:,Beskin:2010:MHDflows:,Meier:TheEngineParadigm,Gra-Jac:2014:MNRAS:,Punsly:2015:BHmag:,Crinquand:2021:PhD:,Kol-Sha-Tur:2023:EPJC:}. As a
starting point, one can consider the electro-vacuum description, in which the
electromagnetic field tensor $F_{\mu\nu}$ satisfies the Maxwell equations
\cite{Mis-Tho-Whe:1973:Gra:}
\beq
 \nabla_{[\lambda} F_{\mu\nu]} = 0, \quad
 \nabla^\mu F_{\mu\nu} =0,
 \label{Eq:Maxwell}
\eeq
where $F_{\alpha \beta} = \partial_\alpha A_\beta - \partial_\beta A_\alpha$ is the electromagnetic tensor.

The Wald solution \cite{Wald:1974:PHYSR4:}, which was obtained by taking advantage of the fact that any linear combination of the Killing vectors generates an exact solution of (\ref{Eq:Maxwell}) \cite{1966AIHPA...4...83P}, is characterized by a vector potential constructed from the timelike and azimuthal Killing vectors, $\xi^\mu_{(t)}$ and $\xi^\mu_{(\phi)}$. For the charged Wald solution, describing a Kerr BH embedded with (test) charge, in an external (test) magnetic field, asymptotically uniform and aligned with the spin axis, the electromagnetic potential is given by
\beq
A^\mu = \frac{B}{2}\,\xi^\mu_{(\phi)}
      + \left(aB - \frac{Q}{2M}\right)\xi^\mu_{(t)},
\label{eq:Awald}
\eeq
where $B$ is the asymptotic magnetic field strength. The nonvanishing covariant components are
\bea
A_t &=& \frac{B}{2}\,g_{t\phi}
      + \left(aB-\frac{Q}{2M}\right) g_{tt}, \label{eq:Awald_t}\\
A_\phi &=& \frac{B}{2}\,g_{\phi\phi}
      + \left(aB-\frac{Q}{2M}\right) g_{t\phi},
\label{eq:Awald_phi}
\eea
with $g_{tt} = -(1-2Mr/\Sigma)$, $g_{t\phi} = -2Mar\sin^2\theta/\Sigma$
and $g_{\phi\phi} = A\sin^2\theta/\Sigma$, where
$\Sigma = r^2 + a^2\cos^2\theta$, $\DD = r^2-2Mr+a^2$ and $A = (r^2+a^2)^2 - a^2\DD\sin^2\theta$ are the metric functions of the Kerr spacetime. The event horizon is located at $r_+ = M + \sqrt{M^{2}-a^{2}}$. Asymptotically, $A_\phi \to \tfrac{1}{2}Br^2\sin^2\theta$ and $A_t \to -Q/r$, so that $Q$ and $B$ have their intended meanings of electric charge and asymptotic magnetic field strength.

We now display the electric and magnetic fields as measured by a locally non-rotating observer, the zero angular momentum observer (ZAMO)
\cite{1970ApJ...162...71B,1972ApJ...178..347B}, whose tetrad is
\bea
e_{\hat{0}} &=& u_{(Z)}, \qquad
e_{\hat{1}} = \sqrt{\DD/\Sigma}\; e_1, \\
e_{\hat{2}} &=& e_2/\sqrt{\Sigma}, \qquad
e_{\hat{3}} = \sqrt{\Sigma/A}\; e_3/\sin\theta,
\eea
where $u^\nu_{(Z)} = \Gamma (1,0,0,\omega)$ is the ZAMO four-velocity as
seen by an observer at rest at infinity, with
$\Gamma = \sqrt{A/(\Sigma \DD)}$ and $\omega = 2 M a r/A$. The field
components measured by the ZAMO are
\beq
E_{\hat{i}} = E_\mu\,e^\mu_{\hphantom{\mu}{\hat i}}, \qquad
B_{\hat{i}} = B_\mu\,e^\mu_{\hphantom{\mu}{\hat i}}.
\eeq
The ZAMO is the natural frame here because it is dragged along with the spacetime, so it measures no spurious electric field arising from
its own rotation. It is also the frame that connects most directly to the fiducial observer (FIDO) used by the GRPIC code (see Sec.~\ref{sec:grpic}).

The unipolar induction effect arises whenever a magnetic field is set into
rotation: in the frame of a static observer, the motion of the field
generates an electric field $\vec{E} = -\vec{v}\times\vec{B}/c$. A
rotating BH immersed in an external magnetic field experiences this situation due to the frame-dragging effect, which forces the magnetic field lines to co-rotate near the horizon. In a plasma-filled magnetosphere, the full electromagnetic problem must be solved self-consistently, because the conductive plasma reacts back on the field: field lines are dragged by the rotating material, poloidal currents generate a toroidal component, and the field lines threading the horizon rotate at an angular velocity $\Omega_{F}$, which in general differs from that of the BH, $\Omega_H = a/(r_+^2+a^2)$. In the force-free limit, $\Omega_{F}\approx\Omega_{H}/2$. As a
consequence, two light surfaces appear on which the corotation velocity
of the field lines equals the speed of light as measured by the ZAMO. The outer light surface is the analog of the pulsar light cylinder: it lies at a cylindrical radius $\approx c/\Omega_{F}$, and beyond it the field lines cannot remain rigidly anchored to the BH and are swept back into a toroidal spiral. The inner light surface lies inside the ergosphere, where the frame-dragging angular velocity $\omega$ exceeds $\Omega_{F}$ so strongly that corotation with the field lines would require superluminal motion relative to the ZAMO. Inside it, the plasma is forced to fall towards the horizon. Rigid corotation of the plasma with the field is therefore possible only between the two light surfaces.

Two scalar invariants of the electromagnetic field are used repeatedly below,
\beq
\mathcal{I}_1 = \tfrac{1}{2}F_{\mu\nu}F^{\mu\nu} = B^2 - E^2,
\quad
\mathcal{I}_2 = \tfrac{1}{4}F_{\mu\nu}{}^{*}\!F^{\mu\nu}
= -\,\vec{E}\cdot\vec{B}.
\label{eq:invariants}
\eeq
Their crucial property is that they are observer-independent: the surface
$\vec{E}\cdot\vec{B}=0$ has the same location for the ZAMO, for a static
observer, and for the FIDO of the simulation. Since only the component of
the electric field parallel to $\vec{B}$ can accelerate a particle along a field line without being screened by the plasma, $\mathcal{I}_2$ is the quantity that determines the sign of the charges moving inward to the horizon, and
its vanishing separates regions of opposite behavior.

\subsection{Horizon potential and Wald's argument}
\label{subsec:potential}

The event horizon of a Kerr BH is a Killing horizon generated by
$\chi^\mu = \xi^\mu_{(t)} + \Omega_H\,\xi^\mu_{(\phi)}$, with the corresponding electrostatic potential
$\Phi = -\chi^\mu A_\mu$. Because $\chi^\mu$ is null and tangent to the generators, $\Phi$ is constant over the horizon. Evaluating $\Phi$ for the charged Wald solution (\ref{eq:Awald}) gives the electrostatic potential difference
between the horizon and infinity
\beq
\Delta\Phi_H = \Phi_H - \Phi_\infty = \frac{Q-2aMB}{2M}
             = \frac{Q - Q_{\rm W}}{2M},
\label{eq:deltaphi}
\eeq
which vanishes for the Wald charge $Q_{\rm W} = 2aMB$. 

The interpretation is straightforward. The energy required to bring a test charge $q$ from rest at infinity to the horizon is $q\,\Delta\Phi_H$. For an uncharged rotating BH, $\Delta\Phi_H = -Q_{\rm W}/2M < 0$, so
positive charges are energetically favored and absorbed preferentially, while negative charges are rejected. The BH charges up, $\Delta\Phi_H$ increases, and the process is self-limiting. Thus, the Wald selective accretion argument drives the system toward $\xi = 1$.

Two caveats motivate everything that follows. First, $\Delta\Phi_H$ is a global quantity, defined between the horizon and infinity. A charged particle in a magnetosphere does not travel along an arbitrary path from infinity, but slides along a magnetic field line, and the potential drop along a given field line does not vanish at $Q = Q_{\rm W}$ \cite{Kom:2022:MNRAS:}. Second, the argument is purely electrostatic and does not account for where on the horizon the charge arrives. The local field structure addresses both objections, to which we now turn.

\subsection{The separatrix $\vec{E}\cdot\vec{B}=0$ and selective accretion}\label{subsec:separatrix}

For a stationary axisymmetric field with $A_\mu = (A_t,0,0,A_\phi)$ and
$\partial_t = \partial_\phi = 0$, we have
\beq
\vec{E}\cdot\vec{B}
 = -\frac{1}{\sqrt{-g}}\,
   \frac{\partial\left(A_t,\,A_\phi\right)}{\partial\left(r,\,\theta\right)}
 = -\frac{\partial_r A_t\,\partial_\theta A_\phi
       - \partial_\theta A_t\,\partial_r A_\phi}{\sqrt{-g}},
\label{eq:EBjacobian}
\eeq
where $\sqrt{-g} = \Sigma\sin\theta$. The
condition $\vec{E}\cdot\vec{B}=0$ is therefore the statement that the
gradients of $A_t$ and $A_\phi$ are parallel, i.e., that the electrostatic
potential is locally a function of the flux function alone, $A_t = A_t(A_\phi)$. This is the familiar degeneracy condition of
force-free electrodynamics \cite{Gra-Jac:2014:MNRAS:}: the vacuum Wald field satisfies it only on a surface, not globally, and it is exactly the failure of degeneracy elsewhere that drives the charging process.

\begin{figure*}
\includegraphics[width=\textwidth]{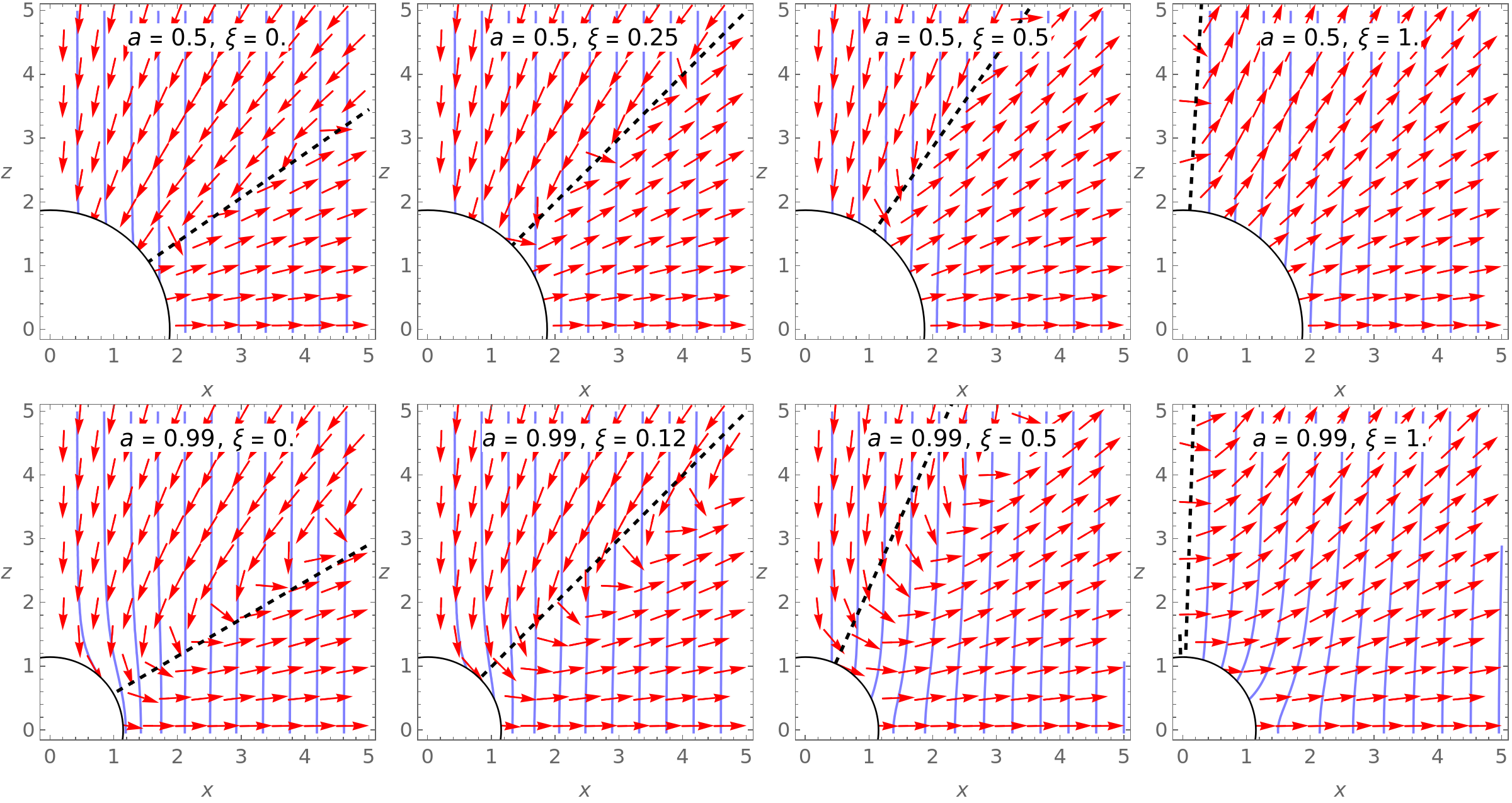}
\caption{Structure of the electric field close to the horizon of a Kerr BH in the Wald solution, illustrating the mechanism of selective accretion. Because the configuration is axisymmetric and there is reflection symmetry about the equatorial plane, only one meridional quadrant is shown, in coordinates $x = r\sin\theta$, $z = r\cos\theta$. Red arrows give the electric field, blue curves the magnetic field lines, and the solid black
arc marks the outer horizon, $r_+$. The dashed curve is the separatrix $\vec{E}\cdot\vec{B} = 0$, Eq.
(\ref{eq:separatrix}): above it, the electric field has a component antiparallel to $\vec{B}$ and drives positive charges along the field
lines towards the BH. Below it, the component is parallel, and positive charges are expelled. The upper row shows a moderately rotating BH,
$a = 0.5\,M$, and the lower row a near-extreme one, $a = 0.99\,M$. In each row, the induced electric charge increases from left to right, from $Q = 0$ to $Q = Q_{\rm W} = 2aMB$. As the BH charges up, the separatrix migrates from the equatorial region towards the rotation axis, progressively shrinking the polar zone in which positive charge is captured \cite{Rue-Ruf:2024:EPJC:}, until at $Q = Q_{\rm W}$ it collapses onto the axis. In the near-extreme case, the migration is considerably faster, and the partial expulsion of the magnetic flux from the horizon (Meissner effect) is visible in the
bending of the field lines.}
\label{fig:selectiveAC}
\end{figure*}

For the charged Wald solution, at first order in the spin parameter, Eq. (\ref{eq:EBjacobian}) leads to the simple result
\beq
\vec{E}\cdot\vec{B}
 = -\frac{B\cos\theta}{r^{2}}
   \left[\,Q - MaB\left(3\cos^{2}\theta - 1\right)\right]
   + \mathcal{O}(a^{2}).
\label{eq:EBslow}
\eeq
Off the equatorial plane, the invariant vanishes on the conical surface
\beq
\xi = \frac{3\cos^{2}\theta_{\rm s} - 1}{2}
\qquad\Longleftrightarrow\qquad
\cos^{2}\theta_{\rm s} = \frac{1+2\xi}{3},
\label{eq:separatrix}
\eeq
which we refer to as the separatrix. 

Beyond the slow-rotation limit, Eq.~(\ref{eq:EBslow}) is no longer adequate and the separatrix must be obtained from the full expression (\ref{eq:EBjacobian}). Evaluated on the horizon, the resulting condition is a quadratic equation for $Q$ \cite{Rue-Ruf:2024:EPJC:}
\beq
\mathcal{C}_2(a,\theta)\,Q^{2} + \mathcal{C}_1(a,\theta)\,Q
 + \mathcal{C}_0(a,\theta) = 0,
\label{eq:quadratic}
\eeq
whose coefficients are lengthy rational functions of $a$ and $\cos^2\theta$ and are most conveniently generated symbolically. Of the two roots, only one connects continuously to the slow-rotation branch (\ref{eq:separatrix}) as $a\to0$. This is the physical root used throughout. The second root is large and negative, $\xi \to -\infty$ as $a\to0$, and is discarded.

Figure~\ref{fig:selectiveAC} shows the electric and magnetic field lines of the Wald solution, as well as the separatrix $\vec{E}\cdot\vec{B} = 0$, for selected values of the BH charge and the BH spin, $a = \{0.5,1\}$. Several features follow. For a neutral BH ($\xi = 0$), the separatrix meets the horizon at $\theta_{\rm s} = \arccos(1/\sqrt{3}) \simeq 54.7^\circ$ (the exact Wald solution gives a value close to $60^\circ$). As the BH charges up, $\theta_{\rm s}$ decreases monotonically: the polar cap in which positive charge is captured shrinks, and at $\xi = 1$ the separatrix collapses onto the rotation axis, so that the capture region disappears altogether. This is the local counterpart of the global statement (\ref{eq:deltaphi}). Note, however, that the electric field is still present.

Physically, the separatrix acts as a valve. Above it (smaller $\theta$), the electric field has an antiparallel component to $\vec{B}$: positive charges are driven along the field lines towards the BH, and negative charges are expelled. Below it (larger $\theta$, towards the equator), the parallel component reverses sign and the roles are interchanged. The horizon is thus divided into a polar patch that absorbs positive charge and an equatorial belt that absorbs negative charge, with $\theta_{\rm s}$ marking the boundary.

\subsection{Equilibrium charge from magnetic flux balance}
\label{subsec:equilibrium}

Because the configuration is stationary and axisymmetric, $A_\phi$ plays the role of a magnetic flux function: the magnetic flux threading the polar cap bounded by the colatitude $\theta$ on a surface of constant $r$
is
\beq
\Psi(r,\theta) = 2\pi A_\phi(r,\theta),
\label{eq:fluxfunction}
\eeq
and poloidal field lines are the level sets of $A_\phi$. Evaluating
(\ref{eq:Awald_phi}) on the outer horizon $r_+ = M + \sqrt{M^2-a^2}$, where $\DD=0$ and $r_+^2+a^2 = 2Mr_+$, yields the compact and exact
expression
\beq
A_\phi(r_+,\theta) =
\frac{r_+\left[\,2MB\left(Mr_+ - a^2\right) + aQ\,\right]\sin^2\theta}
     {r_+^2 + a^2\cos^2\theta}.
\label{eq:Aphi_horizon}
\eeq

\begin{figure*}
\includegraphics[width=\textwidth]{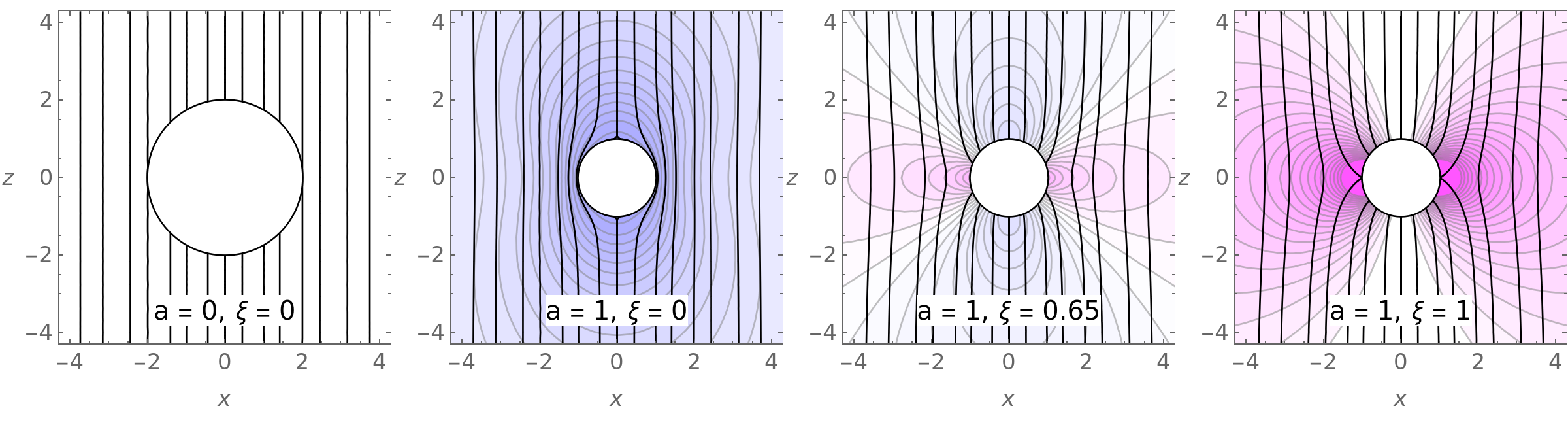}
\caption{Electrostatic potential and magnetic field lines (black curves) of the Wald solution, for selected values of the electric charge $Q$ and spin $a$. The color scale shows the electrostatic potential measured relative to a static observer at infinity, ranging from blue (negative) to magenta (positive). For $Q = a = 0$ (leftmost panel), no potential difference is induced. For $a=M$, the magnetic flux is completely expelled from the horizon at $Q=0$ (Meissner effect). Increasing $Q$ bends the
magnetic field lines back towards the horizon and shifts the potential upwards. For $Q = Q_{\rm W}=2aMB$, the potential vanishes along the symmetry axis \cite{Wald:1974:PHYSR4:}. There is no choice of $Q$ that removes the potential drop along all field lines simultaneously
\cite{Kom:2022:MNRAS:}.}
\label{fig:BHmagneto}
\end{figure*}

For an uncharged BH ($Q=0$), the bracket vanishes identically in the extreme limit $a\to M$ (where $r_+ \to M$): the magnetic flux is completely expelled from the horizon, which is the BH analog of the Meissner effect \cite{Bic-Jan:1985:MNRAS:}. A nonzero charge counteracts this expulsion through the term $aQ$. This effect can be clearly appreciated in Fig.~\ref{fig:BHmagneto}, which compares the magnetic field lines and the electric potential difference of the Wald solution, for selected values of the BH spin $a = \{0,1\}$ and charge $\xi = \{0,0.65,1\}$. As $ Q$ increases, the magnetic field lines are progressively pulled back toward the horizon. We stress that this is a property of the electro-vacuum solution. In a current-carrying magnetosphere, the expulsion is largely suppressed \cite{Kom-McK:2007:MNRAS:}, an issue to which we will return in
Sec.~\ref{sec:summary}.

The BH stops charging when the two patches deliver equal and opposite amounts of charge per unit time. However, this situation does not occur when the two patches have the same surface area. Charged particles do not cross field lines but are tied to them, so only those sliding along lines that intersect the horizon can reach it. The charging capacity of a patch is therefore set by the magnetic flux it subtends, not by its solid angle. Using the flux function (\ref{eq:fluxfunction}), the balance condition reads
\bea
\Psi(r_+,\theta_{\rm s}) = \tfrac{1}{2}\,\Psi(r_+,\pi/2).
\label{eq:balance}
\eea
In the slow-rotation limit, the horizon field is
$B_\perp = B\cos\theta$ on a sphere, so that
$\Psi(\theta)\propto\sin^{2}\theta$ and Eq. (\ref{eq:balance}) gives $\sin^{2}\theta_{\rm s} = 1/2$, i.e., $\theta_{\rm s} = 45^\circ$. Combining this angle with the separatrix relation
(\ref{eq:separatrix}) yields at once $(a\ll M)$
\beq
\xi_{\rm eq} = \frac{3\cdot\tfrac{1}{2}-1}{2} = \frac{1}{4},
\qquad
Q_{\rm eq} = \frac{Q_{\rm W}}{4} = \frac{aMB}{2}.
\label{eq:qquarter}
\eeq

Imposing $\cos^{2}\theta_{\rm s} = 1/2$ in the exact horizon condition (\ref{eq:quadratic}) and retaining the branch continuous with
(\ref{eq:separatrix}), the equilibrium charge is obtained in closed analytic form. Introducing $\sigma \equiv \sqrt{M^{2}-a^{2}} = r_{+}-M$, the result is the remarkably compact expression
\beq
\xi_{\rm eq} = \frac{\sigma\left(5M-\sigma\right)}{4M\left(M+3\sigma\right)} = \frac{\left(r_{+}-M\right)\left(6M-r_{+}\right)}{4M\left(3r_{+}-2M\right)}.
\label{eq:xiclosed}
\eeq
Equation (\ref{eq:xiclosed}) reproduces both limits analytically: for $a \to 0$ one has $\sigma \to M$ and $\xi_{\rm eq} \to 1/4$, recovering Eq.~(\ref{eq:qquarter}), while for $a \to M$ the factor $\sigma$ vanishes
and with it the equilibrium charge, the horizon flux (\ref{eq:Aphi_horizon}) having been expelled. From Eq. (\ref{eq:xiclosed}), we can write the function $Q_{\rm eq}(a)$, which turns out to increase with $a$ until a maximum, $Q_{\rm eq,max} \simeq 0.388\,M^{2}B$, near $a \simeq 0.9\,M$, and then decreases towards zero in the extreme limit. The decline at high spin is a consequence of Eq.~(\ref{eq:Aphi_horizon}): as flux threading the horizon is expelled, the field lines along which charges flow cease to intersect the horizon, and the charging mechanism switches off.

We now anticipate the comparison of the above analytic result for the equilibrium charge with the numerical solutions, shown in Fig.~\ref{fig:chargespin} of Sec. \ref{sec:results}. The analytic result, Eq.~(\ref{eq:xiclosed}), is represented by the black curve, and the value given by the slow-rotation limit, Eq. (\ref{eq:qquarter}), by the black dotted horizontal line.  We emphasize that the black curve has not been adjusted to fit anything: it follows from weighting the two patches by magnetic flux rather than by area, and the flux weighting is the physically appropriate one because particles are guided by $\vec{B}$. 

Equation (\ref{eq:xiclosed}) has been obtained using the fixed separatrix angle of the slow-rotation regime. The direct use of Eq.~(\ref{eq:Aphi_horizon}) into  (\ref{eq:balance}) leads to a closed-form bisection angle independent of both $B$ and $Q$,
\beq
\cos^{2}\theta_{\rm s} = \frac{r_+^{2}}{2r_+^{2} + a^{2}} ,
\label{eq:balanceangle}
\eeq
which indeed leads to $45^\circ$ at $a\to 0$ and in the first-order slow-rotation regime ($r_+\to2M$), but then increases for larger spin values, reaching $\arccos(1/\sqrt{3})\simeq 54.7^\circ$ in the extreme limit ($r_+\to M$). Using (\ref{eq:balanceangle}) instead of the fixed slow-rotation regime value lowers the predicted equilibrium charge at large spin by roughly a factor of two (last two columns of
Table~\ref{tab:analytic}), and moves the peak of $Q$ from $a\simeq0.9$ to
$a\simeq0.75$. Both prescriptions agree in the slow-rotation limit and
both vanish in the extreme case, but the fixed-angle version appears closer to the numerical simulation result at intermediate spins.

\begin{table}[t]
\caption{Analytic equilibrium charge obtained by solving the exact horizon
separatrix condition (\ref{eq:quadratic}) at the flux-balance colatitude.
Columns~2--3 use the fixed value $\theta_{\rm s}=45^\circ$; columns~4--5 use the spin-dependent bisection angle
(\ref{eq:balanceangle}). Charges are in units of $M^{2}B$.}
\label{tab:analytic}
\begin{ruledtabular}
\begin{tabular}{ccccc}
 & \multicolumn{2}{c}{$\theta_{\rm s}=45^\circ$}
 & \multicolumn{2}{c}{$\theta_{\rm s}(a)$, Eq.~(\ref{eq:balanceangle})} \\
$a$ & $\xi_{\rm eq}$ & $Q_{\rm eq}$ & $\xi_{\rm eq}$ & $Q_{\rm eq}$ \\
\hline
0.2  & 0.2500 & 0.100 & 0.2462 & 0.099 \\
0.5  & 0.2488 & 0.249 & 0.2240 & 0.224 \\
0.7  & 0.2435 & 0.341 & 0.1923 & 0.269 \\
0.9  & 0.2155 & 0.388 & 0.1269 & 0.228 \\
0.95 & 0.1889 & 0.359 & 0.0943 & 0.179 \\
0.99 & 0.1204 & 0.238 & 0.0449 & 0.089 \\
\end{tabular}
\end{ruledtabular}
\end{table}


\subsection{Dependence on the magnetic field geometry}
\label{subsec:geometry}

The equilibrium charge derived in Sec.~\ref{subsec:equilibrium} is not a
universal number but a property of the asymptotically uniform Wald field. The criterion itself, Eq.~(\ref{eq:balance}), is geometry-independent, but the angle value depends on how the normal component of the field, $B_{\perp}$, distributes over the horizon. Let us now consider the split-monopole model, obtained by reversing the Blandford-Znajek monopole across the equatorial plane \cite{Bla-Zna:1977:MNRAS:}. An equatorial current sheet sustains it and carries no magnetic charge. This model is widely used as an elementary model of a jet-producing magnetosphere \cite{Kom:2004b:MNRAS:,Ken-etal:2024:PRD:}. Its horizon electromagnetic potential is $A_{\phi}^{\rm SM} = -P|\cos\theta|$, so $B_{\perp} = P/r_{H}^{2}$ is constant over each hemisphere. The enclosed flux scales as
$1-\cos\theta$, and the magnetic-flux horizon bisection condition yields
\beq
\cos\theta_{\rm s} = \tfrac{1}{2},
\qquad \theta_{\rm s} = 60^{\circ}.
\label{eq:60deg}
\eeq
Thus, only in this case do magnetic flux and area bisection yield the same result, Eq. (\ref{eq:60deg}), whereas in the Wald field, the $\cos\theta$ profile concentrates flux towards the pole and shifts the bisector to $45^{\circ}$. The estimate $Q_{\rm eq} = Q_{\rm W}/4$, Eq.~(\ref{eq:qquarter}), is specific to the uniform-field case, and different magnetic-field geometries lead to other values.

This result is astrophysically relevant, since realistic magnetospheres are unlikely to be driven by a single field. The superposition of an internal split monopole with an external uniform field has been shown to produce a paraboloidal, jet-like structure when the two components are aligned and closed, and loop-like structures with magnetic null points when they are anti-aligned \cite{Ken-etal:2024:PRD:}. The generalization to arbitrary inclination, using the Bi\v{c}\'ak-Jani\v{s} solution \cite{Bic-Jan:1985:MNRAS:} for the tilted external component, gives a horizon flux through a polar cap
\beq
\Phi = \pi r_{H}^{2} B_{z} + 2\pi P,
\label{eq:combinedflux}
\eeq
which vanishes identically for $P = -\tfrac{1}{2}r_{H}^{2}B_{z}$
\cite{Zha-etal:2026:PRD:}. Under that condition, the positive polar flux
is exactly canceled by the negative equatorial flux, leaving a radial
null ring on the horizon at $\theta = 60^{\circ}$.

The consequence for the charging problem is direct. Equation (\ref{eq:Aphi_horizon}) shows that the equilibrium charge is controlled by
the flux threading the horizon, and it is the disappearance of that flux
that switches the mechanism off as $a \to M$. Flux cancellation by field
superposition provides an independent and spin-independent route to
the same endpoint: a moderately rotating BH embedded in a sufficiently strong anti-aligned ambient field would deliver no charge to either capture sector and should carry a vanishing induced charge, even far from the extreme BH case. Since the same cancellation has been proposed as a geometric mechanism for jet quenching in compact binaries and for the
absence of a large-scale jet in Sgr~A$^{*}$ \cite{Zha-etal:2026:PRD:},
the induced charge and the jet power are predicted to switch off together.
We stress that these superposition results are obtained for vacuum fields
in the Schwarzschild limit, whereas charging requires rotation, and that
our simulations show the plasma currents modify the vacuum horizon flux
appreciably. A kinetic treatment of the combined geometry is therefore
required to quantify this prediction.

\begin{figure*}
\includegraphics[width=0.45\textwidth]{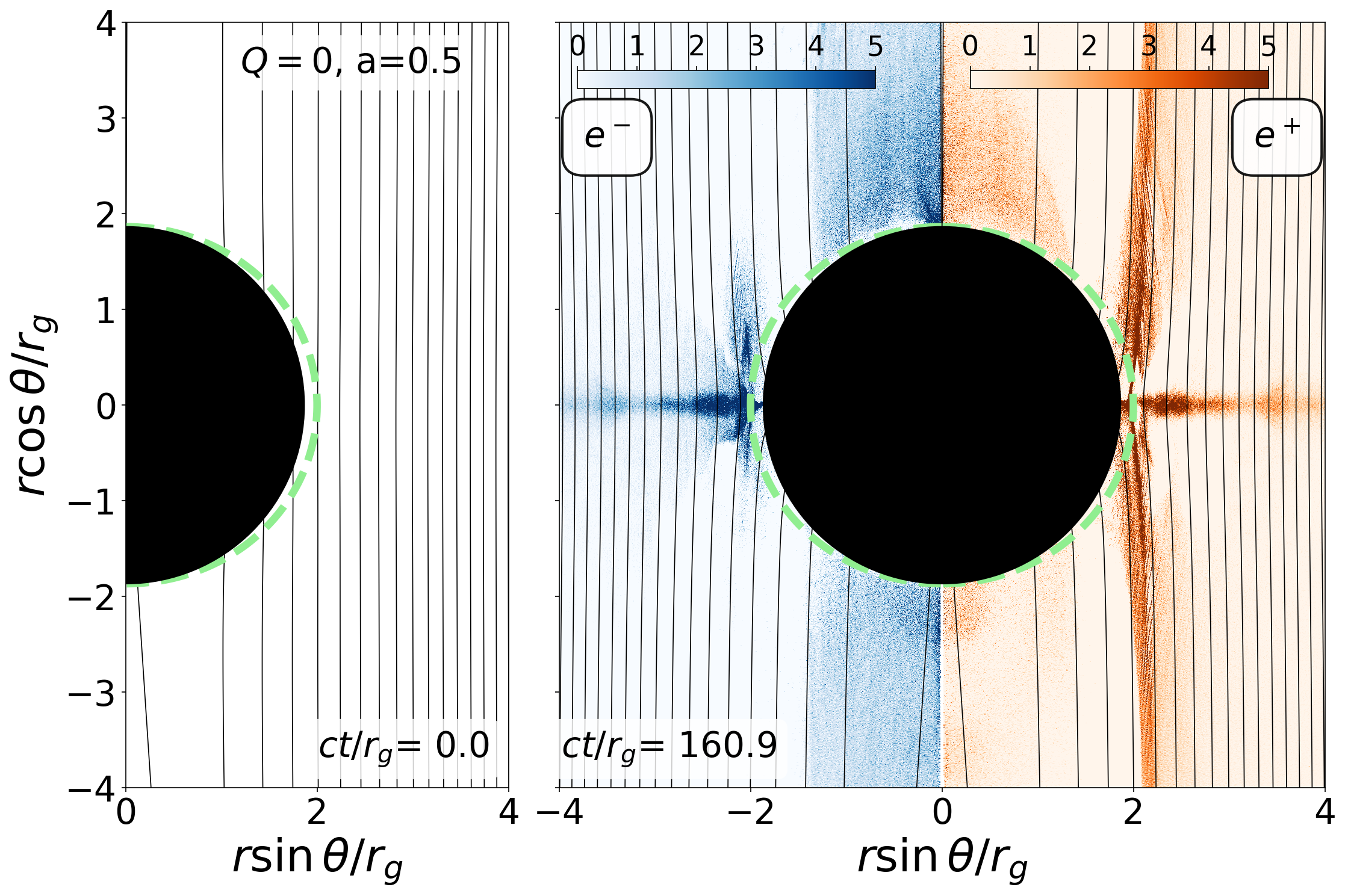}
\includegraphics[width=0.45\textwidth]{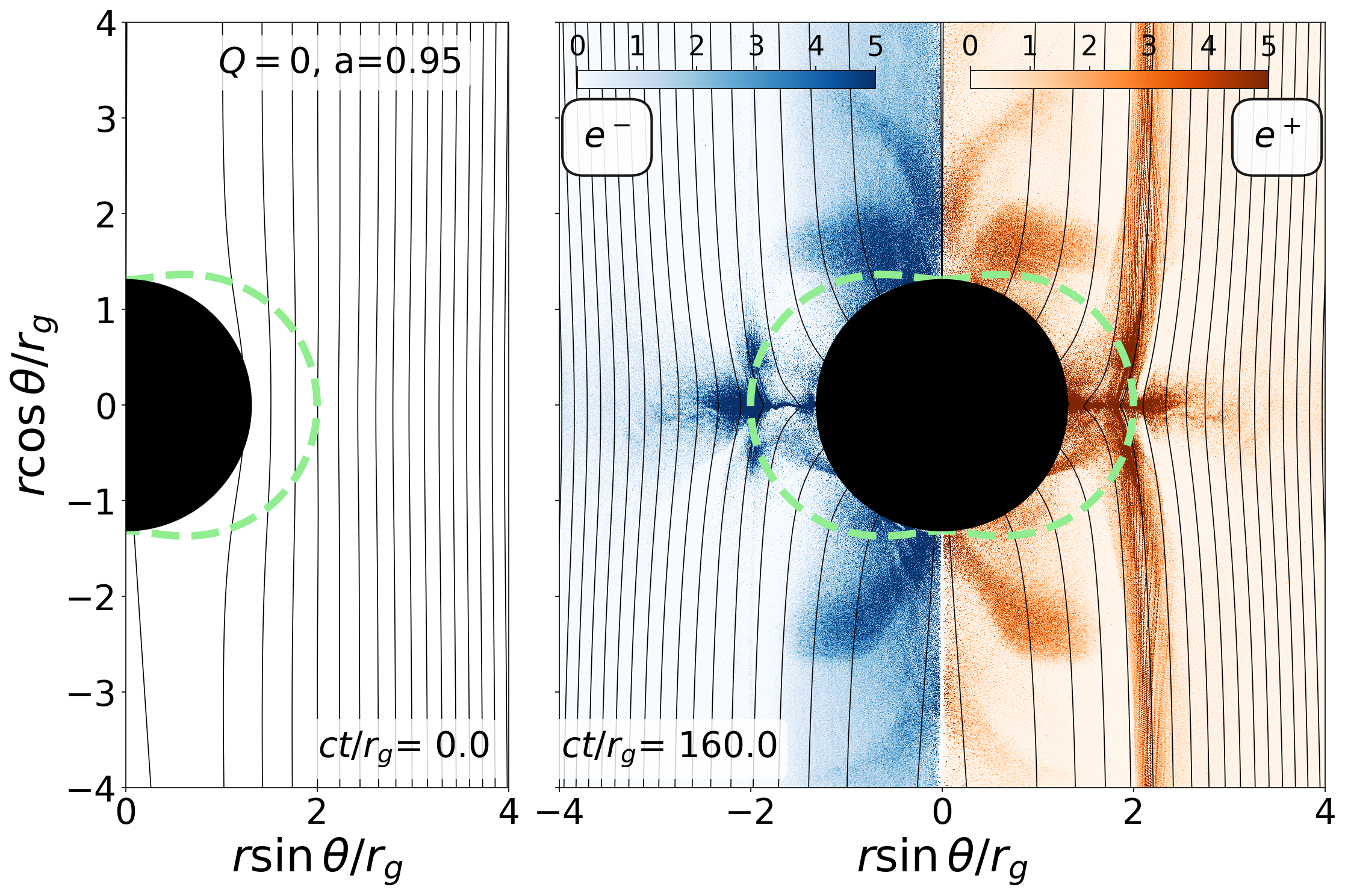}
\includegraphics[width=0.45\textwidth]{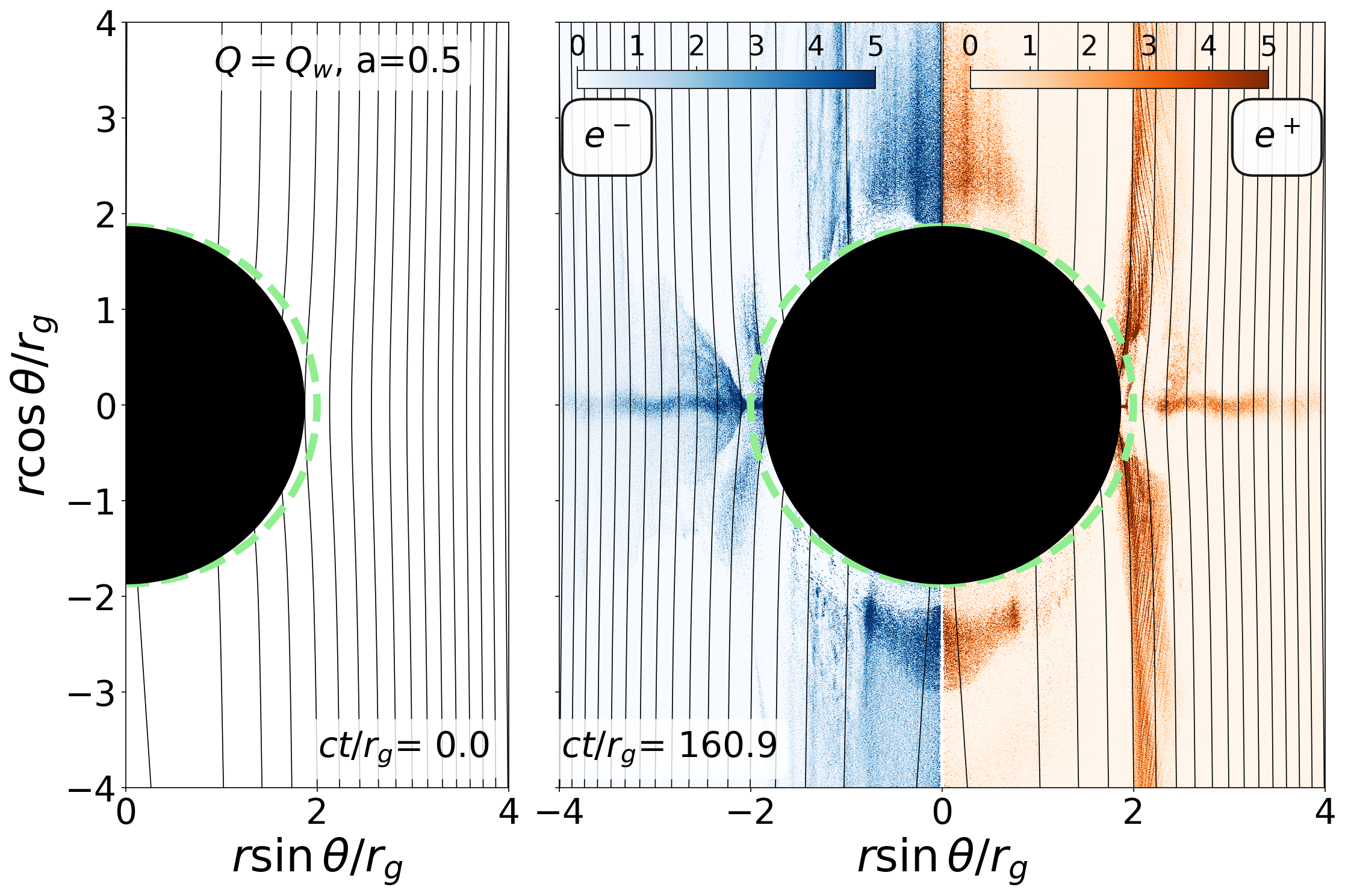}
\includegraphics[width=0.45\textwidth]{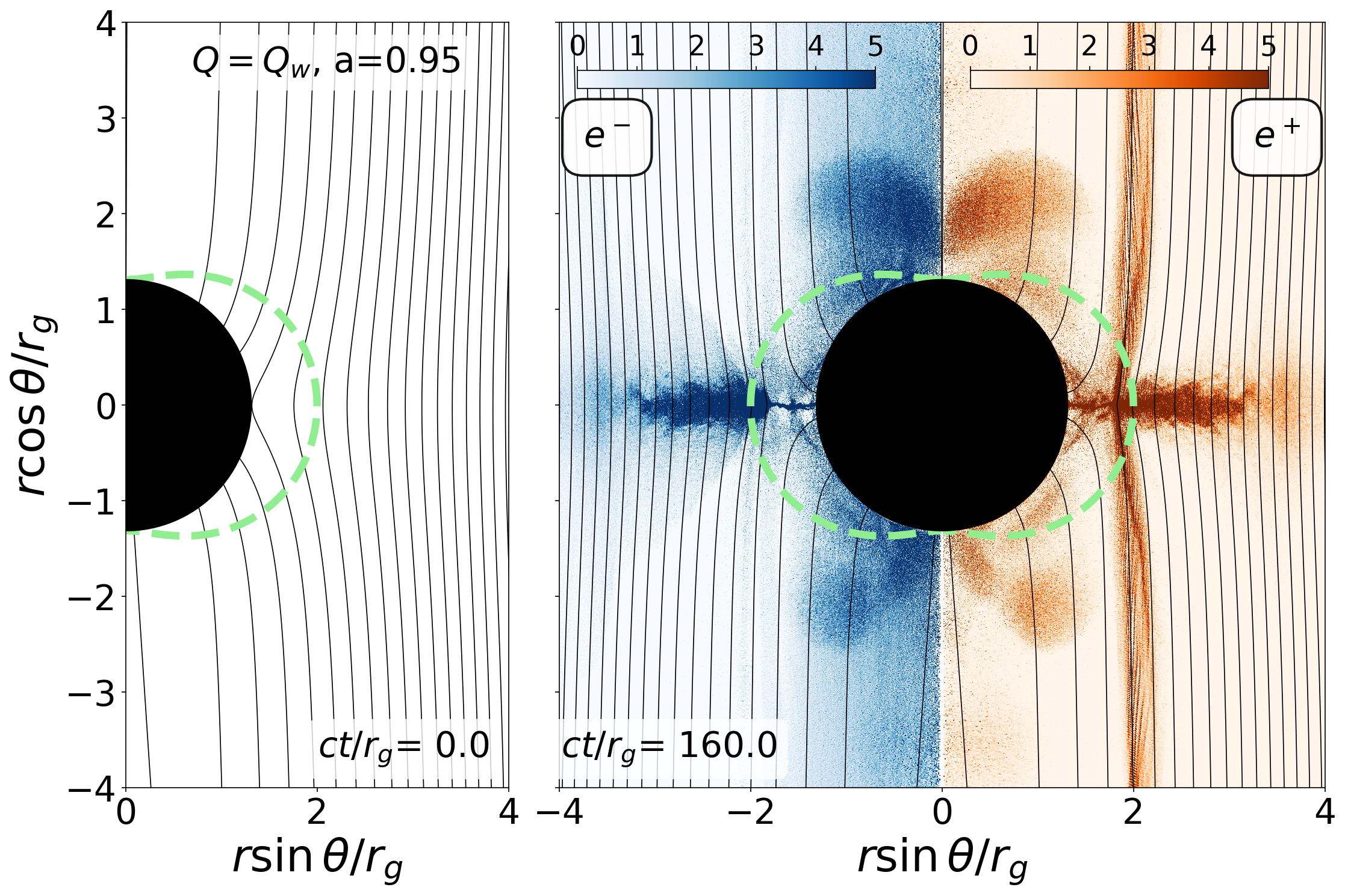}
\caption{Initial and final states of the magnetosphere. In each panel, the
left half shows the initial magnetic field configuration, and the adjacent
full panel shows the final field structure overlaid with the number density normalized to the Goldreich-Julian value, $n/n_{\rm GJ}$, separately for electrons ($e^{-}$, blue) and positrons ($e^{+}$, orange). The left column is $a = 0.5$ and the right column is $a = 0.95$. The top row starts from an uncharged BH ($Q_{0} = 0$), and the bottom row from a Wald-charged BH ($Q_{0} = Q_{\rm W}$). The green dashed line marks the ergosphere. In all
cases, the vacuum solution is replaced within a few $t_{g}$ by a pair-filled,
current-carrying magnetosphere with an equatorial current sheet inside the
ergosphere. The final configurations reached from the two initial charge
states are structurally indistinguishable.}
\label{fig:density}
\end{figure*}
\section{Numerical GRPIC simulations}\label{sec:grpic}

\subsection{The GRPIC method and the GRZeltron code}
\label{subsec:code}

The kinetic approach adopted here is motivated by the nature of the Wald charging problem. Both the force-free and the ideal-MHD descriptions of a BH magnetosphere impose $\vec{E}\cdot\vec{B} = 0$ either exactly or up to an \textit{ad hoc} resistivity, so by construction cannot follow the charge separation that the induced electric field produces, nor the sign-selective capture of particles that ultimately charges the horizon.
Resistive electrodynamics relaxes this constraint \cite{Kom:2004:MNRAS:}, but prescribing a conductivity that is not derived from the plasma itself. In the particle-in-cell (PIC) method, no closure is assumed: the plasma is represented by a large number of \textit{macroparticles} whose trajectories are integrated in the electromagnetic field, which is self-consistently updated, accounting for the currents that those particles deposit on the grid. Non-ideal regions with $\vec{E}\cdot\vec{B}\neq0$, vacuum gaps, collisionless reconnection, and the resulting non-thermal particle distributions emerge from first principles, not introduced by hand.

To perform the simulations, we used the general-relativistic PIC (GRPIC) code GRZeltron, which is the GR extension of the special relativistic PIC code Zeltron \cite{Cer-etal:2013:ApJ:,Cer-Wer:2019:ascl:}. GRZeltron was first
introduced in \cite{Par-Phi-Cer:2019:PRL:}, and a detailed description of its methodology is given in
\cite{Cri-etal:2020:PRL:,Cri-etal:2021:AA:}. The code works in the $3+1$ decomposition of the Kerr spacetime formulated by Komissarov
\cite{Kom:2004:MNRAS:}, using horizon-penetrating Kerr-Schild spherical
coordinates $(t,r,\theta,\varphi)$, which allows placing the inner radial boundary inside the event horizon, without introducing artificial conditions on the horizon itself. Denoting by $\mathbf{B}$ and $\mathbf{D}$ the magnetic and electric fields measured by the fiducial
observer (FIDO), and by $\mathbf{H}$ and $\mathbf{E}$ the auxiliary fields, the Maxwell equations take the following form
\bea
\nabla \cdot \mathbf{D} = &4\pi\rho, \quad
\partial_{t}\mathbf{D} = \nabla\times\mathbf{H} - 4\pi\mathbf{J}, \\
\nabla \cdot \mathbf{B} = &0, \quad
\partial_{t}\mathbf{B} = -\nabla\times\mathbf{E}, \qquad
\label{eq:maxwell31}
\eea
with the constitutive relations
\beq
\mathbf{H} = \alpha\mathbf{B} - \boldsymbol{\beta}\times\mathbf{D},
\qquad
\mathbf{E} = \alpha\mathbf{D} + \boldsymbol{\beta}\times\mathbf{B},
\label{eq:constitutive}
\eeq
where $\alpha$ is the lapse function and $\boldsymbol{\beta}$ the shift
vector. The geometrical content of the metric is carried by $\alpha$, $\boldsymbol{\beta}$ and the spatial metric $h_{ij}$. The field solver retains the standard staggered (Yee-type) mesh and the leapfrog time integration used in the Zeltron code. The current density $\mathbf{J}$ is
obtained by first-order deposition of the macroparticle currents on the
grid. Since this deposition scheme does not conserve charge to machine
precision, the Maxwell-Gauss constraint
$\nabla\cdot\mathbf{D} = 4\pi\rho$ is restored by elliptic divergence
cleaning, i.e., by periodically solving a Poisson equation for a correction potential and adjusting $\mathbf{D}$ accordingly \cite{Cri-etal:2020:PRL:}. This step is essential for the present study, because the quantity we ultimately measure is precisely the Gauss-law flux through the horizon.

Plasma is supplied self-consistently by a radiative pair-production
module: particles inverse-Compton scatter photons of an isotropic
background soft-photon bath, and the resulting high-energy photons
annihilate in flight ($\gamma\gamma\to e^{+}e^{-}$) to create new pairs.
The particle motion accounts for the radiation-reaction force. This
is the same physical setup used to model ergospheric pair discharges in
\cite{Cri-etal:2020:PRL:}, and it guarantees that the plasma density
adjusts dynamically to whatever the electric field requires instead of
being fixed by an injection prescription.

GRZeltron has by now been applied to a broad range of BH-magnetosphere
problems, which provides substantial validation of the numerical machinery
we rely on here: the launching of Blandford-Znajek jets from first
principles \cite{Par-Phi-Cer:2019:PRL:}, multidimensional simulations of
ergospheric pair discharges and gamma-ray emission \cite{Cri-etal:2020:PRL:,Cri-etal:2021:AA:}, the first global 3D GRPIC simulation of a BH magnetosphere together with the corresponding ray-traced synthetic radio images \cite{Cri-etal:2022:PRL:},
magnetospheres magnetically connected to a Keplerian disk with reconnection at the Y-point \cite{ElM-etal:2022:AA:}, reconnection-driven flares and orbiting hot spots as a model for Sgr~A$^{*}$ \cite{ElM-Cer-Cri:2023:AA:}, the kinetic coupling between jet and corona
in accreting BHs \cite{Meh-Cer-Cri:2025:AA:}, and the effect of an inclined external magnetic field on the jet power \cite{Fig-Cer-Par:2025:AA:}. Independent GRPIC codes have reached qualitatively consistent conclusions about the magnetospheric structure
\cite{Bra-Rip-Phi:2021:PRL:}. None of these works, however, tracked the net electric charge accumulated by the BH, which is the specific diagnostic we introduce below.

\subsection{Initial conditions and numerical setup}
\label{subsec:setup}

We use $r_{g} \equiv GM/c^{2}$ as the unit of length and $t_{g} \equiv r_{g}/c$ as the unit
of time. The initial electromagnetic field is initialized with the vacuum Wald
solution (\ref{eq:Awald}). The field strength is set to $B_{0} = r_{g}/r_{L}$, where $r_{L}$
is the electron Larmor radius. This value, which measures the separation between the global and kinetic scales, is many orders of magnitude smaller in any feasible simulation than in astrophysical sources. Still, it respects $r_{L}\ll r_{g}$.

The density required to screen the electric field is given by the
generalized Goldreich-Julian density \cite{Gol-Jul:1969:ApJ:},
\beq
n_{\rm GJ} = \frac{(\Omega_{H} + \Omega_{\rm K})\,B_{0}}{4\pi c e},
\label{eq:ngj}
\eeq
where
\beq
\Omega_{H} = \frac{a}{r_+^{2} + a^{2}}, \qquad
\Omega_{\rm K} = \frac{1}{r_{\rm in}^{3/2} + a}
\label{eq:omegas}
\eeq
are the angular velocity of the BH and the Keplerian angular velocity,
respectively. All plasma densities quoted below are normalized to $n_{\rm GJ}$,
so that $n/n_{\rm GJ}\gtrsim1$ indicates a screened, force-free-like
region and $n/n_{\rm GJ}\ll1$ a charge-starved gap. We set the initial
optical depth of the photons to $\tau_{0} = 60$, which ensures that pair
creation operates, and the magnetosphere is filled even in the lower-spin
runs, where the induced electric field, hence the primary acceleration, is weakest.

The computational grid resolution is $1024\times1024$ cells in the $r$ and
$\theta$ directions, with logarithmic spacing in radius so as to resolve
the near-horizon region while reaching the outer boundary. The inner radial boundary is placed inside the horizon, where the outgoing characteristics guarantee that no information propagates back into the domain. At the outer boundary, we impose outflowing (absorbing) conditions, and at $\theta = 0,\pi$ the axial symmetry conditions $D^{\varphi} = 0$, $B^{\theta} = 0$ and $\partial D^{r}/\partial\theta = 0$
\cite{Cri-etal:2020:PRL:}. Each run employs up to $\approx 8\times10^{6}$ macroparticles and is evolved for $\approx 160\,t_{g}$, i.e., for several tens of horizon rotation periods $4\pi/\Omega_{H}$, long enough for the accumulated charge to reach a saturated plateau (see Fig.~\ref{fig:chargetime}).

The simulations are performed in a 2.5D (axisymmetric) configuration: we
assume $\partial_{\varphi} = 0$ for all quantities, but retain the
three components of the electromagnetic field and particle momenta, so that the toroidal field $B^{\varphi}$ generated by the poloidal currents and the azimuthal drift of the particles are captured. This
is the minimal geometry in which the Blandford-Znajek-like current system
and the ergospheric equatorial current sheet can develop. It also allows us to survey a spin range and run two independent initial charge states for each spin. Non-axisymmetric dynamics is excluded by construction, so the tearing and drift-kink instabilities that fragment the current sheet into orbiting flux ropes in 3D \cite{Cri-etal:2022:PRL:,ElM-Cer-Cri:2023:AA:} cannot be represented. We note, however, that the comparison of 2D and 3D GRZeltron runs in \cite{Cri-etal:2022:PRL:} showed similar global behavior of the magnetosphere. Since the charge we measure is an integral over the whole horizon, it is plausibly insensitive to the sheet's azimuthal structure. Testing this expectation in 3D remains for future work.

For each spin $a = 0.2,\,0.5,\,0.7,\,0.8,\,0.9,\,0.95$ we ran two simulations that differ only in the initial BH charge: neutral ($Q_{0} = 0$), and Wald-charged ($Q_{0} = Q_{\rm W} = 2aMB$). Because these two initial states bracket the range of
charges discussed in the literature, their convergence or divergence constitutes a direct test of whether the saturated charge is an attractor of the kinetic system or merely a memory of the initial data.

\subsection{Measuring the BH charge}
\label{subsec:chargediag}
The accumulated charge on the BH is calculated using Gauss's theorem. In
the $3+1$ formalism, the charge enclosed by a surface of constant radius follows from the volume integral of the Maxwell-Gauss equation, which reduces to a flux integral over that surface
\beq
Q = \frac{1}{4\pi}\int (\nabla\cdot\mathbf{D})\sqrt{h}\,
      \dif r\,\dif\theta\,\dif\varphi
  = \frac{1}{4\pi}\int_{S_{r_{h}}} D^{r}\sqrt{h}\,\dif\theta\,\dif\varphi,
\label{eq:gauss}
\eeq
where $h$ is the determinant of the spatial metric and $r_{h}$ is the horizon radius. In practice, the integral is evaluated on the grid surface closest to $r_{h}$. Since the horizon is enclosed within the computational domain in Kerr-Schild coordinates, this surface is a regular part of the mesh, and no extrapolation is required. Because of the axial symmetry, the $\varphi$ integration is trivial, so that Eq.~(\ref{eq:gauss}) reduces to a one-dimensional quadrature in $\theta$.

We also compute $Q(t)$ by directly time-integrating the radial current flux through the horizon,
\beq
Q(t) = Q(t_{0}) - \int_{t_{0}}^{t} dt'\int_{S_{r_{h}}} J^{r}\sqrt{h}\,\dif\theta\,\dif\varphi,
\label{eq:qcurrent}
\eeq
where $Q(t_{0})$ is fixed by the Gauss-law flux (\ref{eq:gauss}) evaluated on the initial slice. Equation~(\ref{eq:qcurrent}) follows from integrating the continuity equation $\partial_{t}\rho+\nabla\cdot\mathbf{J}=0$ over the volume enclosed by $S_{r_{h}}$; it is therefore not an independent physical measurement of $Q$, but the time-integrated form of the same conservation law that underlies Eq.~(\ref{eq:gauss}). In practice, both quantities were evaluated in every run. All values of $Q$ reported in Sec.~\ref{sec:results} are those obtained from the Gauss-law flux (\ref{eq:gauss}), while Eq.~(\ref{eq:qcurrent}) was used as an internal consistency check on our charge-conserving current deposition scheme. The two agree throughout the saturated phase to within the temporal fluctuations of $Q(t)$, i.e., well within the error bars quoted below.

After an initial transient lasting a few tens of $t_{g}$, during which the magnetosphere fills with pairs and the current sheet forms, $Q(t)$ oscillates about a constant mean value. We therefore define the
equilibrium charge as the time average of $Q(t)$ over the saturated phase,
and quote as its uncertainty the standard deviation over the
same interval (dotted lines and the error bars in Figs.~\ref{fig:chargetime} and \ref{fig:chargespin}).

\section{Results}\label{sec:results}

\subsection{Magnetospheric structure}
\label{subsec:structure}

Figure~\ref{fig:density} shows the initial and final states of the magnetosphere for $a = 0.5$ and $a = 0.95$, for both initial charge states. In every run, the vacuum Wald configuration is modified within a few $t_{g}$. The large unscreened parallel electric field accelerates the seed particles, inverse-Compton scattering produces gamma rays, and pair creation fills the magnetosphere until $n \sim n_{\rm GJ}$ over most of the volume. The poloidal field lines bend back towards the hole under the action of the poloidal currents, a toroidal component $B^{\varphi}$ develops with opposite signs above and below the equator, and an equatorial current sheet forms inside the ergosphere, consistent with previous findings \cite{Par-Phi-Cer:2019:PRL:,Cri-etal:2020:PRL:}. The electron and positron densities are markedly asymmetric near the horizon, which is the direct signature of the charge separation described in Sec.~\ref{subsec:separatrix}: positrons dominate in the polar funnel while electrons accumulate in the equatorial region. At $a = 0.95$, the current sheet is thinner, the plasma supply is more vigorous, and the ergosphere occupies a substantially larger volume. Crucially, the final states reached from $Q_{0} = 0$ and from $Q_{0} = Q_{\rm W}$ are structurally indistinguishable at the same spin, which is the first indication that the initial horizon charge is erased.

\subsection{Convergence to a common saturated charge}
\label{subsec:convergence}

\begin{figure*}
\includegraphics[width=0.85\textwidth]{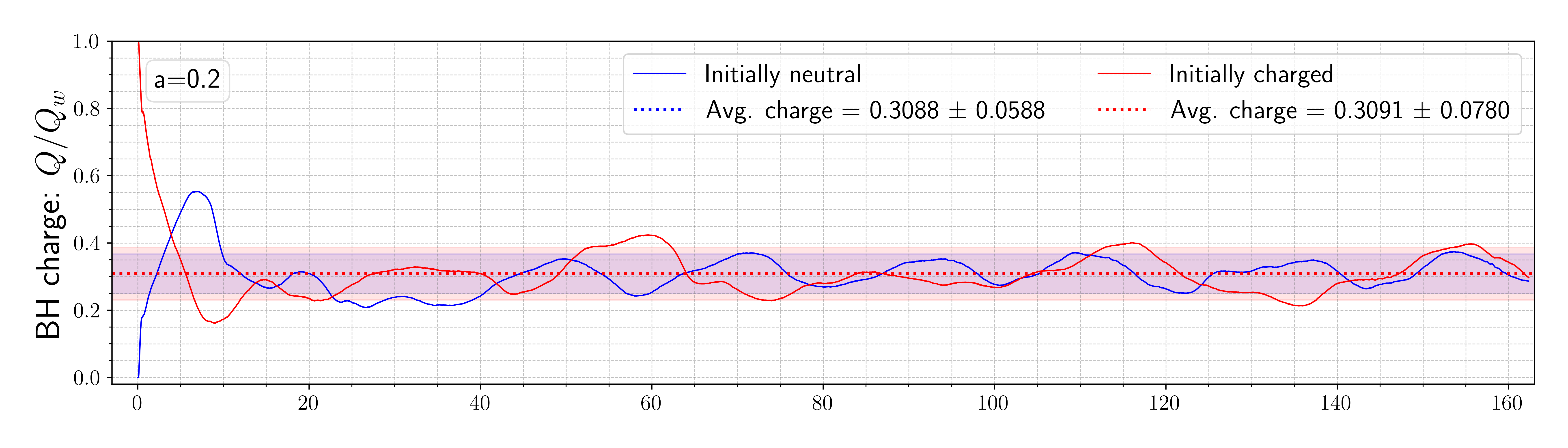}
\includegraphics[width=0.85\textwidth]{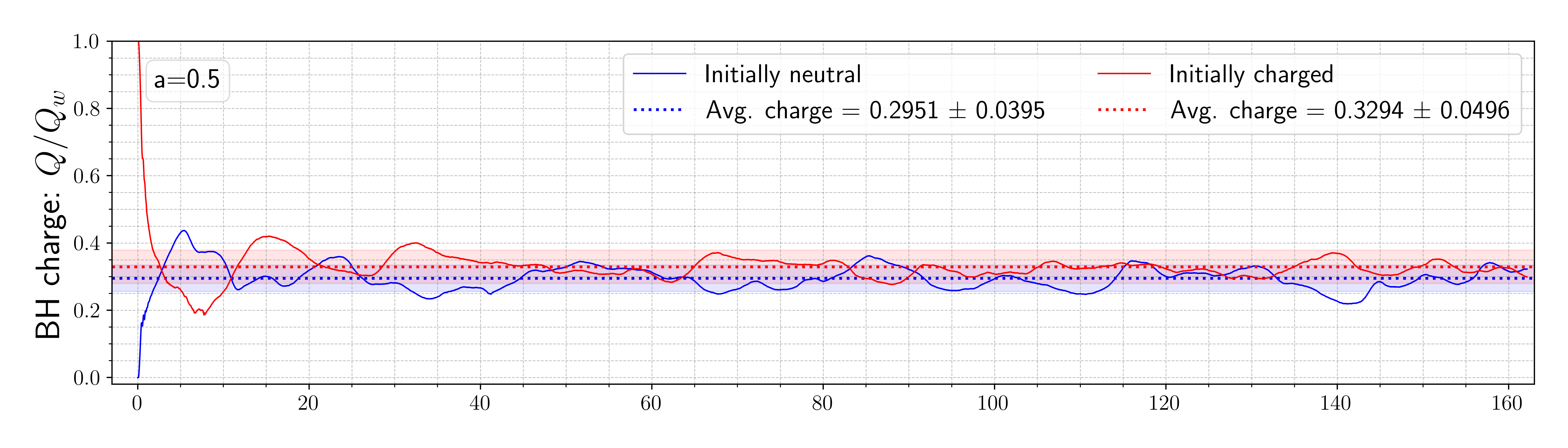}
\includegraphics[width=0.85\textwidth]{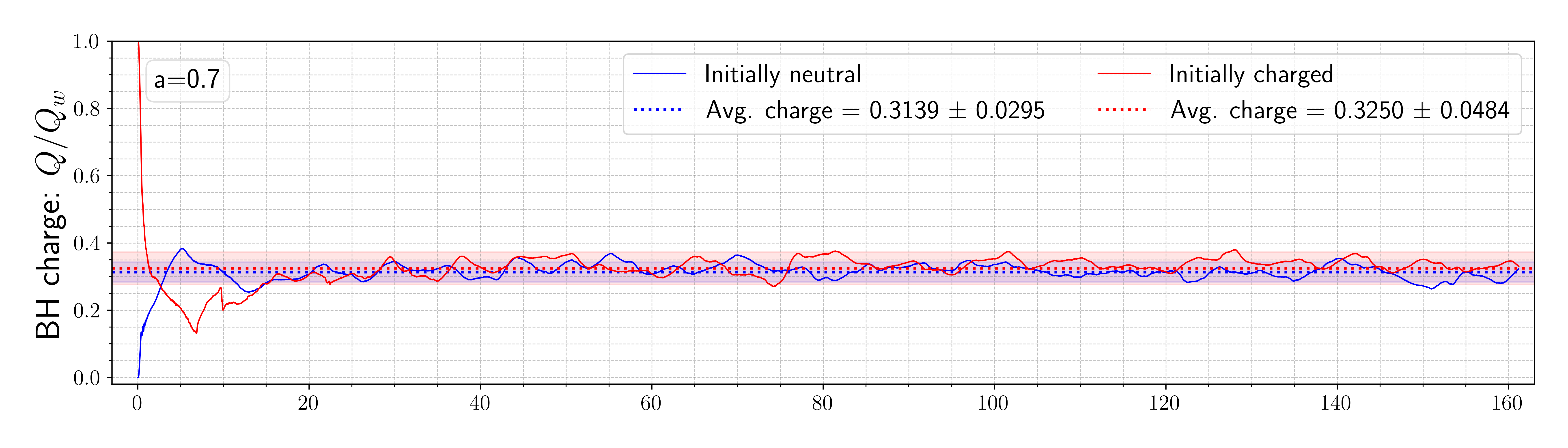}
\includegraphics[width=0.85\textwidth]{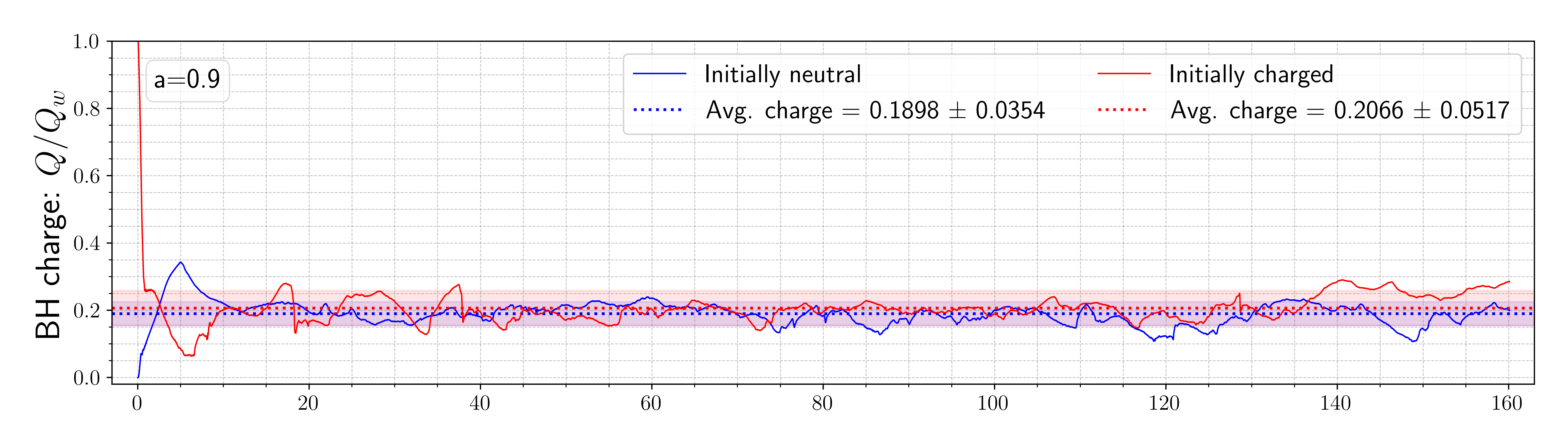}
\includegraphics[width=0.85\textwidth]{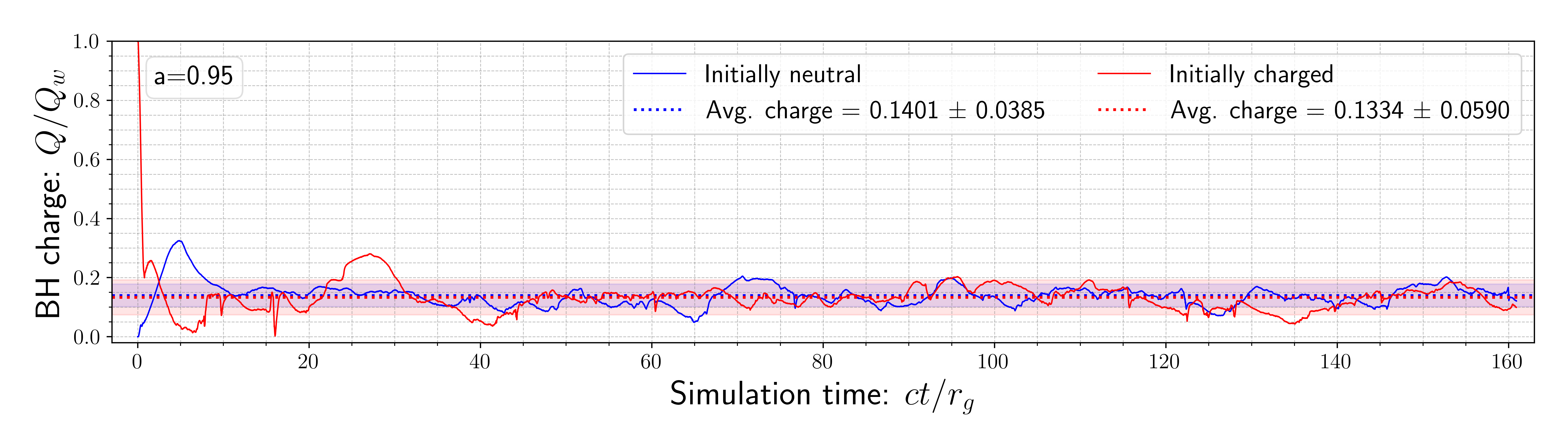}
\caption{Accumulated charge of the BH, normalized to the Wald value, as a function of simulation time $ct/r_{g}$, for spins
$a = 0.2$, $0.5$, $0.7$, $0.9$ and $0.95$ (top to bottom). Blue curves
start from an uncharged BH ($Q_{0} = 0$); red curves start from the Wald
charge ($Q_{0} = Q_{\rm W}$). Dotted horizontal lines are the time averages
over the saturated phase, with the corresponding values and standard
deviations quoted in the legends and in Table~\ref{tab:results}. Although
the two branches begin at opposite extremes and follow entirely different
early histories, they converge within a few tens of $t_{g}$ onto the same
plateau value at every spin.}
\label{fig:chargetime}
\end{figure*}

The central result of this work is displayed in Fig.~\ref{fig:chargetime}. For every spin, the two runs follow completely different early histories: the initially neutral BH charges up rapidly, while the initially Wald-charged BH hole discharges even more rapidly. However, after a transient of a few tens of $t_{g}$, both runs converge onto the same saturated value, which is reached irrespective of the initial charge at every spin (see Table~\ref{tab:results}). 
This result demonstrates that the final charge is a genuine attractor of the kinetic system, not a memory of the initial data. The residual fluctuations about the plateau, of relative amplitude $10$--$30\%$, are physical: they track the intermittent pair discharges and the injection of plasmoids from the equatorial current sheet, which alternately charge and discharge the BH.

The saturated charge is positive in all cases, confirming that the polar capture of positive charge outweighs the equatorial capture of negative charge, and it is well below the Wald value. Averaging the two branches, we obtain $\xi\approx 0.30$--$0.32$ for $a \lesssim 0.7$; $\xi$ then drops to $\xi\approx 0.20$ at $a = 0.9$, and then to $\xi\approx 0.14$ at $a = 0.95$. In no run does the charge approach $\xi = 1$, nor does it approach the much smaller minimum-energy value of \cite{LiXin:2000:PRD:}, which for these spins would be $\xi \lesssim 0.05$.

\begin{table}[t]
\caption{Saturated BH charge $\xi_{\rm eq} = Q/Q_{\rm W}$ measured in the
GRPIC simulations, for runs started from a neutral BH ($Q_{0}=0$) and
from a Wald-charged BH ($Q_{0}=Q_{\rm W}$). Values are time averages
over the saturated phase; uncertainties are the standard deviation of
$Q(t)$ over the same interval.}
\label{tab:results}
\begin{ruledtabular}
\begin{tabular}{cccc}
$a$ & $\xi\ (Q_{0}=0)$ & $\xi\ (Q_{0}=Q_{\rm W})$ & analytic \\
\hline
0.2  & $0.309 \pm 0.059$ & $0.309 \pm 0.078$ & 0.250 \\
0.5  & $0.295 \pm 0.040$ & $0.329 \pm 0.050$ & 0.249 \\
0.7  & $0.314 \pm 0.030$ & $0.325 \pm 0.048$ & 0.243 \\
0.8  & $0.249 \pm 0.031$ & $0.253 \pm 0.048$ & 0.236 \\
0.9  & $0.190 \pm 0.035$ & $0.207 \pm 0.052$ & 0.216 \\
0.95 & $0.140 \pm 0.039$ & $0.133 \pm 0.059$ & 0.189 \\
\end{tabular}
\end{ruledtabular}
\end{table}

\subsection{Spin dependence and comparison with the analytic model}
\label{subsec:spindep}

\begin{figure*}
\includegraphics[width=0.49\textwidth]{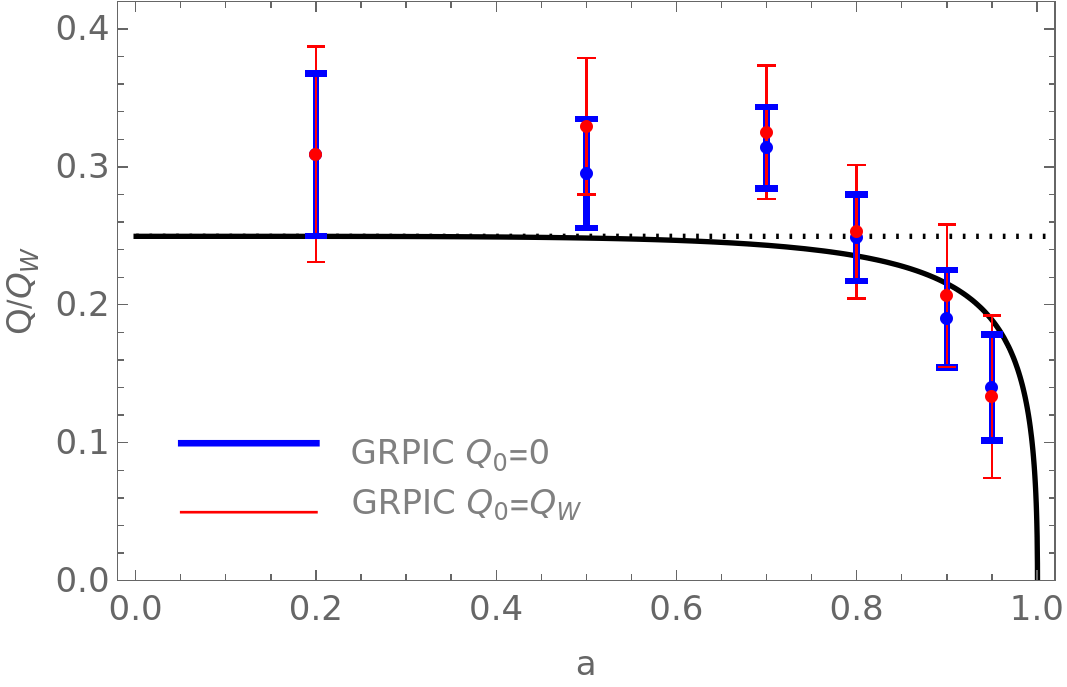}
\includegraphics[width=0.49\textwidth]{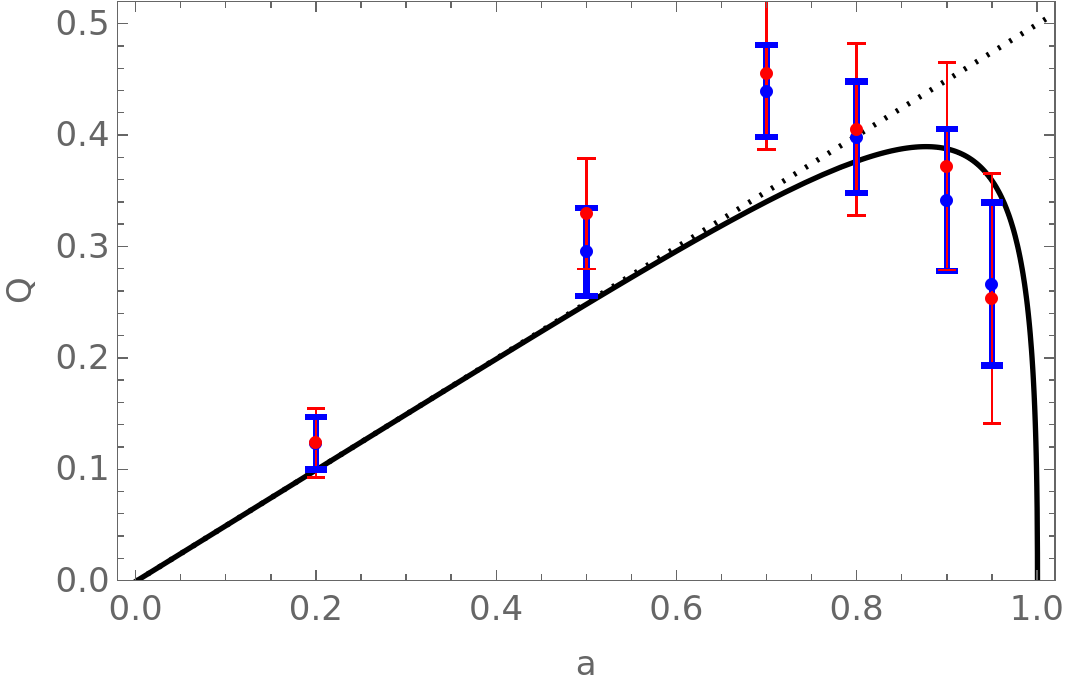}
\caption{Equilibrium charge of a black hole immersed in an external
uniform magnetic field as a function of spin $a$, from our GRPIC
simulations and compared with the analytic prediction. The left panel
gives the charge normalized to the Wald value,
$\xi = Q/Q_{\rm W} = Q/(2aMB)$, the right panel the charge itself in units
of $M^{2}B$. Blue markers denote runs started from an uncharged hole,
$Q_{0} = 0$, red markers runs started from the Wald charge,
$Q_{0} = Q_{\rm W}$; error bars indicate the spread of the charge over the
saturated phase. That the two sets agree within their uncertainties at
every spin shows that the final state is a genuine attractor, reached
irrespective of the initial charge. The solid black curve is the analytic
equilibrium charge of Sec.~\ref{subsec:equilibrium}, obtained by requiring
that the separatrix $\vec{E}\cdot\vec{B} = 0$ bisect the {magnetic}
flux threading one hemisphere of the horizon, Eq.~(\ref{eq:balance}), so
that the polar and equatorial sectors deliver equal and opposite amounts
of charge per unit time. In the slow-rotation limit this criterion places
the separatrix at $\theta_{\rm s} = 45^{\circ}$ and gives
$Q = Q_{\rm W}/4$, shown as the dotted line; the solid curve departs from
it above $a \simeq 0.6$ and falls to zero as $a \to M$, where the
expulsion of magnetic flux from the horizon (Meissner effect) suppresses
the charging mechanism altogether. The simulations reproduce this
behaviour, including the decline towards extremality, although the
measured charges lie systematically somewhat above the analytic curve at
intermediate spins.}
\label{fig:chargespin}
\end{figure*}

Figure~\ref{fig:chargespin} collects the saturated charges as a function of spin and compares them with the analytic prediction of Sec.~\ref{subsec:equilibrium}. Two features stand out. First, in units of the Wald charge, the equilibrium is roughly flat at $\xi_{\rm eq}\approx{0.3}$ for $a\lesssim{0.7}$ and then declines steeply, consistent with the decline of the analytic curve driven by the expulsion of horizon-threading magnetic flux. In absolute units, the charge peaks near $a \approx 0.9$, and falls towards zero in the extreme BH case. Second, the measured charges lie systematically above the analytic curve at intermediate spins, by roughly $20$--$30\%$, while agreeing with it within the error bars at $a=0.9$ and $a=0.95$.

The analytical description in Sec.~\ref{subsec:equilibrium} of the charge equilibrium value has been obtained using the electro-vacuum Wald field only. The simulated magnetosphere, instead, carries currents that modify both the geometry of the poloidal field near the horizon and the location of the $\vec E \cdot \vec B = 0$ separatrix. In particular, the currents suppress the Meissner expulsion \cite{Kom-McK:2007:MNRAS:}, so a higher flux threads the horizon at high spin. Therefore, the polar capture channel remains open for somewhat longer, pushing the equilibrium charge upward relative to the Wald-field estimate. The analytic model also treats the two capture sectors as perfectly efficient, ignoring the finite residence time of particles in the gap and the fact that a fraction of the pairs created near the inner light surface is ejected rather than absorbed. The Wald-field estimate reproduces the low-spin value $\xi\approx{1/4}$ and the decrease towards the extreme case, suggesting that it captures correctly the essential physics driving the system to charge equilibrium.

\section{Summary and conclusions}\label{sec:summary}

We have addressed the Wald charging problem, i.e., whether a rotating BH immersed in an external magnetic field charges up to $Q_{\rm W} = 2aMB$, by performing first-principles axisymmetric GRPIC simulations including self-consistent pair creation and radiation reaction. For every BH spin $a = 0.2$, $0.5$, $0.7$, $0.8$, $0.9$ and $0.95$, we evolved the same asymptotically uniform Wald field from two opposite initial states of the horizon charge, $Q_{0} = 0$ and $Q_{0} = Q_{\rm W}$, and measured the accumulated charge both from the Gauss law flux at the horizon and from the cumulative net number of particles absorbed. Our principal conclusions follow.

(i) The horizon charge is a dynamical attractor. Irrespective of whether the BH starts neutral or Wald-charged, the system relaxes within a few tens of $t_{g}$ to the same positive, saturated charge, and the two branches agree within their uncertainties at every spin. The charge acquired by an astrophysical BH is therefore selected by kinetic plasma processes (e.g., pair creation, current-sheet dynamics, and the sign-dependent capture of particles streaming along the field lines), and not by the initial data or by vacuum electrostatics alone.

(ii) The equilibrium is strongly sub-Wald. The measured charge is $\xi\approx{0.30}$ for $a\lesssim{0.7}$, falling to $\xi \approx 0.20$ at $a=0.9$ and $\xi \approx 0.14$ at $a=0.95$. In no case does the BH reach $\xi=1$. The Wald charge thus behaves as an upper limit, in agreement with the expectation of \cite{Kom:2022:MNRAS:} and in tension with the complete-screening scenario of \cite{Kin-Pri:2021:ApJL:}. The measured values also exceed by a factor of a few the minimum-energy charge of \cite{LiXin:2000:PRD:}. Thus, neither of the two extremes proposed in the literature is realized.

(iii) A simple physical criterion accounts for the magnitude. Requiring the separatrix $\vec{E}\cdot\vec{B} = 0$ to bisect the magnetic flux threading one hemisphere of the horizon places it at $\theta_{\rm s} = 45^{\circ}$ in the slow-rotation limit, and predicts $Q = Q_{\rm W}/4$, within $20$--$30\%$ of the measured plateau. Applying the same criterion to the exact Wald field reproduces the decline at high spin. We stress that it is the magnetic and not the electric flux that must be balanced: for $Q_{0} = 0$ the ingoing and outgoing electric fluxes are equal by Gauss law, so an electric-flux criterion would predict no evolution at all. The magnetic-flux criterion encodes the kinematic fact that particles reach the horizon only by sliding along field lines.

(iv) Extremality and the Meissner effect are related but distinct. In units of the Wald charge, the equilibrium decreases monotonically with spin, whereas in absolute units it peaks near $a \approx 0.9\,M$ and falls to almost zero as $a \to M$. In vacuum, this limit coincides with the complete expulsion of magnetic flux from the horizon. In the GRPIC magnetosphere, the currents present suppress that expulsion \cite{Kom-McK:2007:MNRAS:}, so the two effects need not coincide: in the plasma-filled case, the vanishing of the charge at extremality and the vanishing of the horizon-threading flux have distinct physical origins.

Astrophysically, the persistence of a small but non-vanishing charge matters precisely because it is small. For $\xi < 1$, the horizon-infinity potential difference is not screened away: the accelerating gap near the inner light surface survives, pair discharges continue to load the magnetosphere, and the Blandford-Znajek mechanism is not quenched by charging. At the same time, the residual charge modifies the separatrix geometry and hence the polar and equatorial capture sectors, and thus the injection of plasma that feeds the jet. The equilibrium charge which enters in analyses of the magnetic Penrose process and the acceleration of ultra-high-energy cosmic rays \cite{Tur-etal:2020:ApJ:}, as well as in constraints on the charge of Sgr~A$^{*}$ \cite{Zaj-etal:2018:MNRAS:}, can now be compared against a kinetically determined value.

We close by stating limitations of the present treatment. Our simulations are axisymmetric, so non-axisymmetric instabilities of the equatorial current sheet, tilted-field configurations, and any $m \neq 0$ redistribution of charge are excluded by construction. The scale separation $r_{g}/r_{L}$ is many orders of
magnitude smaller than in any real source, and the pair multiplicity is controlled by a single parameter $\tau_{0}$. Full 3D calculations, a broader survey in $\tau_{0}$ and in the field inclination, and a treatment
that incorporates the current-modified field geometry into the analytic separatrix condition are the natural next steps.

\begin{acknowledgments}
This work was partially supported by the Czech Science Foundation Grant (GA\v{C}R) No.~\mbox{23-07043S}, the internal grants \mbox{IGS/21/2026}, \mbox{IGS/27/2026}  and \mbox{SGS/24/2024} from the Silesian University in Opava. This project has received funding from the European Research Council (ERC) under the European Union’s Horizon 2020 research and innovation program (Grant Agreement No. 863412). The authors acknowledge the use of the Laniakea Computing Cluster at the Silesian University in Opava in carrying out this research. 
A. Tursunov is co-funded by the European Union under the Marie Skłodowska-Curie Actions (Physics for Future – Grant Agreement No. 101081515). FA acknowledges MSK25 for its support as part of the project "Support for science and research in the Moravian-Silesian Region 2025".

\end{acknowledgments}

\bibliographystyle{unsrt}
\bibliography{reference}

\end{document}